\documentclass[final,3p,times]{elsarticle}

\biboptions{numbers,sort&compress}
\usepackage[figuresright]{rotating}
\usepackage{graphicx, subcaption, color, epsfig, dsfont, amssymb, amsmath, amsthm, amsfonts, amsbsy, mathrsfs, mathtools, delarray}
\usepackage[colorlinks, linkcolor=green]{hyperref} 

\newtheorem{theorem}{Theorem}
\newtheorem{definition}{Definition}

\newtheorem{algorithm}{Algorithm}

\journal{Peer Review}

\begin{document}

\begin{frontmatter}




\title{Data-based approximate policy iteration for nonlinear continuous-time optimal control design \tnoteref{note}}
\tnotetext[note]{\color{red}This is the updated version of the submitted paper (with the same title). If the reviewers can find this version, you can review this version instead. Thanks.}

\author[beihang]{Biao Luo} \ead{biao.luo@hotmail.com}
\author[beihang]{Huai-Ning Wu} \ead{whn@buaa.edu.cn}
\author[qatar]{Tingwen Huang} \ead{tingwen.huang@qatar.tamu.edu}
\author[Chinese_Academy]{Derong Liu} \ead{derong.liu@ia.ac.cn}

\address[beihang]{Science and Technology on Aircraft Control Laboratory, Beihang University (Beijing University of Aeronautics and Astronautics), Beijing 100191, P. R. China}
\address[qatar]{Texas A\& M University at Qatar, PO Box 23874, Doha, Qatar}
\address[Chinese_Academy]{State Key Laboratory of Management and Control for Complex Systems, Institute of Automation, Chinese Academy of Sciences, Beijing 100190, P. R. China}

\begin{abstract}
This paper addresses the model-free nonlinear optimal problem with generalized cost functional, and a data-based reinforcement learning technique is developed. It is known that the nonlinear optimal control problem relies on the solution of the Hamilton-Jacobi-Bellman (HJB) equation, which is a nonlinear partial differential equation that is generally impossible to be solved analytically. Even worse, most of practical systems are too complicated to establish their accurate mathematical model. To overcome these difficulties, we propose a data-based approximate policy iteration (API) method by using real system data rather than system model. Firstly, a model-free policy iteration algorithm is derived for constrained optimal control problem and its convergence is proved, which can learn the solution of HJB equation and optimal control policy without requiring any knowledge of system mathematical model. The implementation of the algorithm is based on the thought of actor-critic structure, where actor and critic neural networks (NNs) are employed to approximate the control policy and cost function, respectively. To update the weights of actor and critic NNs, a least-square approach is developed based on the method of weighted residuals. The whole data-based API method includes two parts, where the first part is implemented online to collect real system information, and the second part is conducting offline policy iteration to learn the solution of HJB equation and the control policy. Then, the data-based API algorithm is simplified for solving unconstrained optimal control problem of nonlinear and linear systems. Finally, we test the efficiency of the data-based API control design method on a simple nonlinear system, and further apply it to a rotational/translational actuator system. The simulation results demonstrate the effectiveness of the proposed method. 
\end{abstract}


\end{frontmatter}

\section{Introduction} \label{Sec_1}
The nonlinear optimal control problem has been widely studied in the past few decades, and a large number of theoretical results \cite{lewis2013optimal,bertsekas2005dynamic, hull2003optimal} have been reported. However, the main bottleneck for their practical application is that the so-called Hamilton-Jacobi-Bellman (HJB) equation should be solved. The HJB equation is a first order nonlinear partial differential equation (PDE), which is difficult or impossible to solve, and may not have global analytic solutions even in simple cases. For linear systems, the HJB equation results in an algebraic Riccati equation (ARE). In 1968, Kleinman \cite{kleinman1968iterative} proposed a famous iterative scheme for solving the ARE, where it was converted to a sequence of linear Lyapunov matrix equations. In \cite{saridis1979approximation}, the thought of the iterative scheme was extended to solve HJB equation, which was successively approximated by a series of generalized HJB (GHJB) equations that are linear Lyapunov function equations (LFEs). To solve the GHJB equation, Beard et al. \cite{beard1997galerkin,beard1998approximate} proposed a Galerkin approximation approach where a detailed convergence analysis was provided. By using neural network (NN) for function approximation, the iterative scheme was further extended to constrained input systems \cite{abu2005nearly} and discrete-time systems \cite{chen2008generalized}. However, most of these approaches are model-based which require the accurate mathematical model of the system.

With the fast developments of science technologies, many industrial systems (such as systems in aeronautics and astronautics, chemical engineering, mechanical engineering, electronics, electric power, traffic and transportation) become more and more complicated due to their large scale and complex manufacturing techniques, equipments and procedures. One of the most prominent features for these systems is the presence of vast volume of data accompanied by the lack of an effective process physical model that can support control design. Moreover, the accurate modelling and identification of these systems are extremely costly or impossible to conduct. On the other hand, with the extensive applications of digital sensor technologies, and the availability of cheaper measurement and computing equipments, more and more system information could be extracted for direct control design. Thus, the development of data-based control approaches for practical systems is a promising, but still challenging research area.

Over the past few decades, the thoughts of reinforcement learning (RL) techniques have been introduced to study the optimal control problems \cite{lewis2013reinforcement, lendaris2008higher, lewis2012reinforcement}. RL is a machine learning technique that has been wildly studied from the computational intelligence and machine learning scope in the artificial intelligence community \cite{sutton1998reinforcement, bertsekas1996neuro, hafner2011reinforcement, kaelbling1996reinforcement}. RL technique refers to an actor or agent that interacts with its environment and aims to learn the optimal actions, or control policies, by observing their responses from the environment. As one of the most popular RL schemes, approximate/adaptive dynamic programming (ADP) \cite{bertsekas1996neuro, powell2007approximate} uses value function approximation structures (such as, linear or nonlinear function approximation \cite{tsitsiklis1997analysis}) for the implementation of the RL algorithms. ADP solves the dynamic programming problem forward-in-time, and thus avoids the so-called ``curse-of-dimensionality". Moreover, RL and ADP methods have the ability to find an optimal control policy from unknown environment, which makes RL a promising method for data-based control design. In \cite{sutton1998reinforcement}, Sutton and Barto suggested a definition of RL method, i.e., any method that is well suited to solve RL problem can be considered to be a RL method, where the RL problem is defined in terms of optimal control of discrete-time Markov decision processes. This obviously established the relationship between the RL method and optimal control problem. Especially for discrete-time systems\cite{ni2013adaptive, he2005reinforcement, liu2013policy, he2007reinforcement, al2008discrete, lu2008direct, yang2008control, zhang2009neural, wang2011adaptive, lewis2011reinforcement}, the thoughts of RL and ADP have been introduced for optimal control design with known or unknown system models. For example, heuristic dynamic programming (HDP) was used to solve the optimal control problem of nonlinear discrete-time systems \cite{al2008discrete}, or with control constraints \cite{zhang2009neural} or time delays \cite{zhang2011optimal}; Inspired by the action dependent HDP (ADHDP), Si and Wang \cite{si2001online} developed a direct HDP (DHDP) approach for online learning an optimal control policy; Lewis and Vamvoudakis \cite{lewis2011reinforcement}  derived two ADP algorithms for linear system: output feedback policy iteration and value iteration, which only require measurements of input/output data. Fu et al. \cite{fu2011adaptive} investigated the adaptive learning and control for multiple-input-multiple-output system based on ADP; A finite-horizon optimal control problem was studied in \cite{wang2011adaptive} by introducing a $ \varepsilon $-error bound; And finite-time problem with control constraint was considered by proposing a dual heuristic programming (DHP) scheme \cite{heydari2013finite} with single NN; To involve the effects of NN approximation errors, a neural HDP method in \cite{yang2012reinforcement} was applied to learn state and output feedback adaptive critic control policy of nonlinear discrete-time affine systems with disturbances; Dierks \& Jagannathan \cite{dierks2012online} proposed a time-based ADP, which is an online control method without using value and policy iterations; Globalized DHP algorithms \cite{liu2013iterative, wang2012optimal, liu2012neural} were developed by using three NNs for estimating system dynamics, cost function and its derivatives, and control policy, where model NN construction error was considered. 

RL is considerably more difficult for continuous-time systems than for discrete-time systems, and fewer results are available \cite{lewis2012reinforcement}. Doya \cite{doya2000reinforcement} introduced using appropriate approximators for estimating value function to minimize the temporal difference error in RL approach; Murray et al. \cite{murray2002adaptive} suggested two policy iteration algorithms that avoid the necessity of knowing the internal system dynamics either by evaluating the infinite horizon cost associated with a control policy along the entire stable state trajectory, or by using measurements of the state derivatives to form the Lyapunov equations; Vrabie et al. \cite{vrabie2009adaptive} extended their result and proposed a new policy iteration algorithm to solve the linear quadratic regulator (LQR) problem online along a single state trajectory; A nonlinear version of this algorithm was presented in \cite{vrabie2009neural} by using a NN approximator; Vamvoudakis and Lewis \cite{vamvoudakis2010online} gave an online policy iteration algorithm which tunes synchronously the weights of both actor and critic NNs for the nonlinear optimal control problem; In \cite{liu2013decentralized}, ADP was employed to design stabilizing control strategy for a class of continuous-time nonlinear interconnected large-scale systems.  But those methods are partially model-based \cite{murray2002adaptive, vrabie2009adaptive, vrabie2009neural, modares2013integral} or completely model-based \cite{vamvoudakis2010online, liu2013decentralized}. Recently, some data-based RL methods have been reported. For example, data-based policy iteration \cite{Jiang2012computational} and Q-learning \cite{lee2012integral} algorithms were developed for linear systems; The nonlinear optimal control problem was considered in \cite{zhang2011data, modaresadaptive2013adaptive}, but they require a prior model identification procedure and then model-based adaptive methods were used. To the best of our knowledge, the problem of model-free RL method design for nonlinear continuous-time optimal control problem is still an open problem, which motivates the present study.

In this paper, we consider the general optimal control problem of continuous-time nonlinear systems with completely unknown model, and develop a model-free approximate policy iteration (API) method for learning the optimal control policy from real system data. The rest of the paper is arranged as follows. The problem description and some preliminary results are presented in Sections \ref{Sec_2} and \ref{Sec_3}. Then, data-based API methods are developed for constrained and unconstrained optimal control problems in Sections \ref{Sec_4} and \ref{Sec_5} respectively.  Finally, the effectiveness of data-based API method is tested in Section \ref{Sec_6}, and a brief conclusion is given in Section \ref{Sec_7}.

\textit{Notation}: $\mathbb{R}, \mathbb{R}^n$ and $\mathbb{R}^{n\times m} $  are the set of real numbers, the $ n $-dimensional Euclidean space and the set of all real   matrices, respectively.  $ \Vert \cdot \Vert $ denotes the vector norm or matrix norm in $ \mathbb{R}^n$ or $\mathbb{R}^{n\times m} $   , respectively. The superscript $ T $ is used for the transpose and  $ I $ denotes the identify matrix of appropriate dimension. $ \bigtriangledown \triangleq \partial / \partial x $  denotes a gradient operator notation. For a symmetric matrix $ M, M>(\geq) 0$  means that it is a positive (semi-positive) definite matrix.  $ \Vert v \Vert ^2_M \triangleq v^T M v $  for some real vector  $ v $ and symmetric matrix    $ M>(\geq) 0$  with appropriate dimensions. $ C^1(\Omega) $  is a function space on $ \Omega $  with first derivatives are continuous.   Let $ \Omega $ and $\mathcal{U} $  be compact sets, denote $ \mathcal {D}  \triangleq \lbrace (x,u) | x \in \Omega, u \in \mathcal{U} \rbrace $. For column vector functions $ s_1(x,u) $ and $ s_2(x,u) $  , where $ (x,u) \in \mathcal {D} $ ， define inner product $ \langle s_1(x,u), s_2(x,u)\rangle_{\mathcal {D} } \triangleq \int_{\mathcal {D} } s_1^T(x,u) s_2(x,u) d(x,u) $  and norm  $ \Vert s_1(x,u) \Vert _{\mathcal {D} } \triangleq \left( \int_{\mathcal {D} } s_1^T(x,u) s_1(x,u) d(x,u) \right)^{1/2}$ .

\section{Problem description} \label{Sec_2}
Let us consider the following continuous-time nonlinear system:
\begin{equation}\label{eq_2.1}
\dot{x}(t) = f(x(t)) + g(x(t))u(t), x(0)=x_0 
\end{equation}
where $[x_1~ ... ~x_n]^T \in \Omega \subset \mathbb{R}^n$ is the state, $x_0$ is the initial state and $u = [u_1~ ... ~u_m]^T \in \mathcal{U} \subset \mathbb{R}^m$ is the control input. Assume that, $ f(x) + g(x)u(x) $ is Lipschitz continuous on a set $\Omega$  that contains the origin, $ f(0)=0 $, and that the system is stabilizable on  $\Omega$, i.e., there exists a continuous control function $ u(x) $  such that the system is asymptotically stable on  $\Omega$. $ f(x) $  and $ g(x) $  are continuous vector or matrix functions of appropriate dimension, the accurate models of which are assumed to be \textit{unknown} in this paper.

The optimal control problem under consideration is to find a state feedback control law $ u(t) = u(x(t)) $  such that the system \eqref{eq_2.1} is closed-loop asymptotically stable, and minimize the following generalized infinite horizon cost functional:
\begin{equation}\label{eq_2.2}
V(x_0) \triangleq \int_{0}^{+\infty}(Q(x(t))+W(u(t)))dt
\end{equation}
where $ Q(x) $ and $ W(u) $ are positive definite functions, i.e., for $ \forall x \neq 0, u \neq 0, Q(x)>0, W(u)>0 $, and $ Q(x)=0, W(u)=0  $ only when $ x = 0, u = 0 $. Then, the optimal control problem is briefly presented as
\begin{equation} \label{eq_2.3}
u(t) = u^*(x) \triangleq \arg\min_u V(x_0)
\end{equation}

\section{Preliminary works} \label{Sec_3}
In this section, some related work will be presented. Before starting, the definition of admissible control \cite{beard1997galerkin,abu2005nearly} is given.

\begin{definition} \label{def_3.1}
(Admissible control) For the given system \eqref{eq_2.1}, $ x \in \Omega $, a control $ u(x) $ is defined to be admissible with respect to cost function \eqref{eq_2.2} on $ \Omega $, denoted by  $ u(x) \in \mathfrak{U}(\Omega) $, if, 1)  $ u $ is continuous on $ \Omega $, 2)  $ u(0)=0 $, 3)  $ u(x) $ stabilizes the system, and 4)  $ V(x)<\infty, \forall x \in \Omega $. $ \square $ 
\end{definition}

For $ \forall u(x) \in \mathfrak{U}(\Omega) $, its cost function $ V(x) $ of \eqref{eq_2.2} satisfies the following Lyapunov function equation (LFE) \cite{abu2005nearly}:
\begin{equation} \label{eq_3.1}
[\nabla V(x)]^T (f(x) + g(x)u(x) ) +Q(x) + W(u) =0
\end{equation}
where $ V(x) \in C^1(\Omega), V(x) \geq 0 $ and $ V(0) = 0 $. 
From the optimal control theory \cite{lewis2013optimal,bertsekas2005dynamic, anderson2007optimal}, if using the optimal control $ u^*(x) $, the LFE \eqref{eq_3.1} results in the HJB equation
\begin{equation} \label{eq_3.2}
[\nabla V^*(x)]^T (f(x) + g(x)u^*(x) ) +Q(x) + W(u^*) =0.
\end{equation}

\subsection{Constrained optimal control} \label{Sec_3.1} 
For the system \eqref{eq_2.1} with input constraints $ | u_i | \leqslant \beta $, the following nonquadratic form $ W(u) $ for the cost functional \eqref{eq_2.2} can be used \cite{lyashevskiy1996constrained, abu2005nearly, modaresadaptive2013adaptive}:   
\begin{equation}\label{eq_3.3}
W(u) 
= 2\int_{0}^{u} (\phi^{-1} (\mu))^T R d\mu
=2\sum_{l=1}^{m} r_l \int_{0}^{u_l} \phi^{-1} (\mu) d\mu 
\end{equation}
where $\mu \in \mathbb{R}^m, \phi(\cdot) $ is a continuous one-to-one bounded function satisfying $ |\phi(\cdot)| \leqslant \beta $ with $ \phi(0) = 0 $.  Moreover, $ \phi(\cdot) $ is a monotonic odd function and its derivative is bounded. An example of $\phi(\cdot)$ is the hyperbolic tangent $ \tanh(\cdot)$ and $R=diag(r_1~...~r_m)>0$ is a diagonal matrix for simplicity. From \cite{abu2005nearly}, the HJB equation \eqref{eq_3.2} of the constrained optimal control problem is given by
\begin{equation} \label{eq_3.4}
[\nabla V^*]^T \left( f-g \phi(\frac{1}{2} R^{-1} g^T \nabla V^*)\right) + Q(x) 
+2\int_{0}^{-\phi \left( \frac{1}{2} R^{-1} g^T \nabla V^* \right)} (\phi^{-1} (\mu))^T R d\mu = 0 . 
\end{equation}
By solving the HJB equation for $ V^*(x) $, the optimal control policy is obtained with
 \begin{equation}\label{eq_3.5}
 u^*(x) = -\phi \left( \frac{1}{2} R^{-1} g^T(x) \nabla V^* (x) \right).
\end{equation}

For description simplicity, define
\begin{equation}\label{eq_3.6}
 \nu^*(x) \triangleq -\frac{1}{2} R^{-1} g^T(x) \nabla V^* (x)
\end{equation}
then, the HJB equation \eqref{eq_3.4} and optimal control \eqref{eq_3.5} can be briefly rewritten as:
 \begin{equation}\label{eq_3.7}
(\nabla V^*)^T \left(f+g \phi(\nu^*)\right) + Q+2\int_{0}^{\phi ( \nu^*)} (\phi^{-1} (\mu))^T R d\mu =0 
\end{equation}
 \begin{equation}\label{eq_3.8}
u^*=\phi (\nu^*).
\end{equation}
In \cite{abu2005nearly}, the HJB equation \eqref{eq_3.7} is successively approximated with a sequence of LFEs
\begin{equation}\label{eq_3.9}
[\nabla V^{(i+1)}]^T (f + gu^{(i)} ) +Q +2\int_{0}^{u^{(i)}} (\phi^{-1} (\mu))^T R d\mu =0;
 i = 0,1,... 
\end{equation}
where
\begin{equation}\label{eq_3.10}
u^{(i)}=\phi (\nu^{(i)}).
\end{equation}
with
\begin{equation}\label{eq_3.11}
\nu^{(i)} \triangleq -\frac{1}{2} R^{-1} g^T \nabla V^{(i)}.
\end{equation}
By providing an initial control policy $ u^{(0)} \in \mathfrak{U}(\Omega) $, it has been proven in \cite{abu2005nearly} that the solution of the iterative LFE \eqref{eq_3.9} will converge to the solution of the HJB equation \eqref{eq_3.7}, i.e., $\lim_{i \to \infty} V^{(i)} = V^* $ and thus $\lim_{i \to \infty} u^{(i)} = u^* $.
\subsection{Unconstrained optimal control} \label{Sec_3.2}
For the system \eqref{eq_2.1} without input constraints, $ W(u) $ in the cost functional \eqref{eq_2.2} can be selected as a simple quadratic form $ W(u) =\Vert u \Vert ^2_R $ with $ R>0 $. Then, for unconstrained optimal control problem, the HJB equation \eqref{eq_3.2} is written as 
\begin{equation} \label{eq_3.12}
[\nabla V^*(x)]^T f(x) + Q(x) - \frac{1}{4} [\nabla V^*(x)]^T  g(x) R^{-1} g^T(x) \nabla V^*(x) = 0.
\end{equation} 
 and the associated optimal controller is given by 
 \begin{equation}\label{eq_3.13}
u^*(x) = -\frac{1}{2} R^{-1} g^T(x) \nabla V^*(x).
\end{equation} 
In \cite{saridis1979approximation}, the HJB equation \eqref{eq_3.12} was successively approximated by a sequence of LFEs as follows:
\begin{equation}\label{eq_3.14}
[\nabla V^{(i+1)}]^T (f + gu^{(i)} ) +Q(x) + \Vert u^{(i)} \Vert _R^2 =0;
 i = 0,1,... 
\end{equation}
with
\begin{equation}\label{eq_3.15}
u^{(i)} = -\frac{1}{2} R^{-1} g^T(x) \nabla V^{(i)} (x).
\end{equation}
For giving an initial control policy $ u^{(0)} \in \mathfrak{U}(\Omega) $, the convergence of iterative equation \eqref{eq_3.14} with \eqref{eq_3.15} is proved in \cite{saridis1979approximation}.

\noindent \textbf{Remark 1.} It is worth pointed out that the LFEs \eqref{eq_3.9} and \eqref{eq_3.14} are specific forms of the general LFE \eqref{eq_3.1} with different choices of $ W(u) $, where $ V^{(i+1)} (x) $ is the cost function of control policy $ u^{(i)} (x) $. Note that LFE is a linear partial difference equation that is much simpler than the HJB equation. In \cite{beard1997galerkin} and \cite{abu2005nearly}, the LFEs \eqref{eq_3.14} and \eqref{eq_3.9} were solved with Galerkin approximation  and NN methods, respectively. However, these approaches are completely model-based, where system models $ f(x) $ and $ g(x) $ should be accurately known. $ \square $

\section{Data-based approximate policy iteration for constrained optimal control} \label{Sec_4}
In this section, data-based approximate policy iteration (API) method is developed to solve the constrained optimal control problem of system \eqref{eq_2.1}. Since the mathematical model of system dynamics $ f(x) $ and $ g(x) $  are completely unknown, the explicit expression of the associated HJB equation \eqref{eq_3.7} is unavailable. Thus, it is impossible to obtain the solution of HJB equation with model-based approaches. To overcome this problem, we propose a data-based API algorithm to learn the solution of the HJB equation \eqref{eq_3.7} by using the online information of real system rather than system model.
\subsection{Derivation of data-based policy iteration} \label{Sec_4.1}
To derive the data-based API algorithm, we rewrite the system \eqref{eq_2.1} as
\begin{equation}\label{eq_4.1}
\dot{x} = f + gu^{(i)}  + g[u - u^{(i)}]
\end{equation}
for $ \forall u\in \mathcal{U}$. Let us consider  $ V^{(i+1)}(x) $ be the solution of the LFE \eqref{eq_3.9}. By using \eqref{eq_3.9}-\eqref{eq_3.11}, we take derivative of $ V^{(i+1)}(x) $  with respect to time along the state of system \eqref{eq_4.1}
\begin{flalign}\label{eq_4.2}
\dfrac{dV^{(i+1)}(x)}{dt}  & = [\nabla V^{(i+1)}]^T(f + gu^{(i)}) - [\nabla V^{(i+1)}]^Tg[u^{(i)} - u] \nonumber\\
&= -Q - 2\int_{0}^{u^{(i)}} (\phi^{-1} (\mu))^T R d\mu + 2(\nu^{(i+1)})^T R [u^{(i)} - u]\nonumber\\
&= -Q - 2\int_{0}^{\phi(\nu^{(i)})} (\phi^{-1} (\mu))^T R d\mu + 2(\nu^{(i+1)})^T R [\phi(\nu^{(i)}) - u]
\end{flalign}
Integrating both sides of \eqref{eq_4.2} on the interval $ [t, t+\Delta t] $  and rearranging terms yields, 
\begin{flalign}\label{eq_4.3}
2\int_{t}^{t+\Delta t} [\nu^{(i+1)}(x(\tau))]^T R [\phi(\nu^{(i)}(x(\tau))) - u(\tau)] d \tau 
+V^{(i+1)}(x(t)) - V^{(i+1)}(x(t+\Delta t)) \nonumber\\
= \int_{t}^{t+\Delta t} \left(Q(x(\tau)) + 2\int_{0}^{\phi(\nu^{(i)}(x(\tau)))} (\phi^{-1} (\mu))^T R d\mu \right)d\tau
\end{flalign}
In \eqref{eq_4.3}, $ V^{(i+1)}(x) $  and $ \nu^{(i+1)}(x) $  are unknown function and function vector needed to be solved. Given an initial admissible control policy  $ u^{(0)} $, the problem of solving the LFE \eqref{eq_3.9} for  $ V^{(i+1)}(x) $, is transformed to the problem of solving the equation \eqref{eq_4.3} for $ V^{(i+1)} (x)$  and $ \nu^{(i+1)}(x) $. Compared with LFE \eqref{eq_3.9}, equation \eqref{eq_4.3} does not require the explicit mathematical model of system \eqref{eq_2.1}, i.e., $ f(x) $  and $ g(x) $. 

\noindent \textbf{Remark 2.} Note that in iterative equation \eqref{eq_4.3}, the system dynamic models   $ f(x) $  and $ g(x) $ are not required. In fact, their information is embedded in the online measurement of the state  $ x $ and control signal $ u $. Thus, the lack of information about system model does not have any impact on the model-free policy iteration algorithm for learning the solution of HJB equation and the optimal control policy. The resulting control policy learns with the real process behavior, and thus does not suffer from the problem of model inaccuracy or simplifications in the model-based approaches. Furthermore, in contrast to control methods based on the nonparametric identification models, the issue about collecting system data is also incorporated within the learning process and can be concentrated on regions important to the control application. $\square$

\noindent \textbf{Remark 3.} It is noted that the data-based policy iteration with \eqref{eq_4.3} is an ``off-policy" learning method \cite{precup2001off}, which means that the cost function  $ V^{(i+1)} (x)$ of control policy $ u^{(i)}(x) $  can be evaluated by using system data generated with other different control policies $ u $ . Off-policy learning, the ability for an agent to learn about a policy other than the one it is following, is a key element of reinforcement learning. The obvious advantage of off-policy learning is that it can learn the cost function and control policy from states and actions that are selected according to a more exploratory or even random policy. $ \square $

The convergence of the data-based policy iteration with \eqref{eq_4.3} is established in Theorem \ref{theorem_4.1}.

\begin{theorem}\label{theorem_4.1}
Let $ V^{(i+1)}(x) \in C^1(\Omega), V^{(i+1)}(x) \geq 0,  V^{(i+1)}(0) = 0 $ and $ \phi(\nu^{(i+1)}(x)) \in \mathfrak{U}(\Omega) $. $ (V^{(i+1)}(x), \nu^{(i+1)}(x)) $ is the solution of equation \eqref{eq_4.3} iff $($ if and only if $ ) $ it is the solution of the LFE \eqref{eq_3.9} and \eqref{eq_3.11}, i.e., equation \eqref{eq_4.3} is equivalent to the LFE \eqref{eq_3.9} with \eqref{eq_3.11}.
\end{theorem} 

\noindent \textbf{Proof.} From the derivation of equation \eqref{eq_4.3}, it is concluded that if $ (V^{(i+1)}, \nu^{(i+1)})$  is the solution of the LFE \eqref{eq_3.9} with \eqref{eq_3.11}, then $ (V^{(i+1)}, \nu^{(i+1)})$ also satisfies equation \eqref{eq_4.3}. To complete the proof, we have to show that $ (V^{(i+1)}, \nu^{(i+1)})$ is the unique solution of equation \eqref{eq_4.3}. The proof is by contradiction.

Before starting the contradiction proof, we derive a simply fact. Consider
\begin{flalign} \label{eq_4.4}
\lim\limits_{\Delta t \to 0} \frac{1}{\Delta t} \int_{t}^{t+\Delta t} \hbar (\tau) d \tau & =
\lim\limits_{\Delta t \to 0} \frac{1}{\Delta t} \left( \int_{0}^{t+\Delta t} \hbar (\tau) d \tau - \int_{0}^{t} \hbar (\tau) d \tau \right) \nonumber \\
& = \frac{d}{dt} \int_{0}^{t} \hbar (\tau) d \tau \nonumber \\
& =  \hbar (t).
\end{flalign}

From \eqref{eq_4.3}, we have
\begin{flalign}\label{eq_4.5}
\frac{dV^{(i+1)}(x)}{dt} 
&= \lim\limits_{\Delta t \to 0} \frac{1}{\Delta t} \left( V^{(i+1)}(x(t + \Delta t)) - V^{(i+1)}(x(t)) \right) \nonumber\ \\ 
& = 2 \lim\limits_{\Delta t \to 0} \int_{t}^{t+\Delta t} [\nu^{(i+1)}(x(\tau))]^T R [\phi(\nu^{(i)}(x(\tau))) - u(\tau)] d \tau \nonumber \\
& \quad - \lim\limits_{\Delta t \to 0} \frac{1}{\Delta t} \int_{t}^{t+\Delta t} \left(Q(x(\tau)) + 2\int_{0}^{\phi(\nu^{(i)}(x(\tau)))} (\phi^{-1} (\mu))^T R d\mu \right)d\tau.
\end{flalign}
By using the fact \eqref{eq_4.4}, the equation \eqref{eq_4.5} is rewritten as
\begin{flalign}\label{eq_4.6}
\frac{dV^{(i+1)}(x)}{dt} = 2[\nu^{(i+1)}(x(t))]^T R [\phi(\nu^{(i)}(x(t))) - u(t)]
- Q(x(t)) - 2\int_{0}^{\phi(\nu^{(i)}(x(t)))} (\phi^{-1} (\mu))^T R d\mu.
\end{flalign}
Suppose that $ (W(x), \upsilon(x))$ is another solution of equation \eqref{eq_4.3}, where $ W(x) \in C^1(\Omega) $ with boundary condition $ W(0) = 0 $ and $ \phi(\upsilon(x)) \in \mathfrak{U}(\Omega)  $. Thus, $ (W, \upsilon)$ also satisfies equation \eqref{eq_4.6}, i.e.,
\begin{flalign}\label{eq_4.7}
\frac{dW(x)}{dt} = 2\upsilon^T(x(t)) R [\phi(\nu^{(i)}(x(t))) - u(t)]
- Q(x(t)) - 2\int_{0}^{\phi(\nu^{(i)}(x(t)))} (\phi^{-1} (\mu))^T R d\mu.
\end{flalign}
Substituting equation \eqref{eq_4.7} from \eqref{eq_4.6} yields,
\begin{equation} \label{eq_4.8}
\frac{d}{dt} \left(V^{(i+1)}(x) -W(x) \right) = 2[\nu^{(i+1)}(x(t)) - \upsilon(x(t))]^T R [\phi(\nu^{(i)}(x(t))) - u(t)]. 
\end{equation}
This means that equation \eqref{eq_4.8} holds for $ \forall u \in \mathcal{U}$. If letting $ u = \phi(\nu^{(i)})$,  we have
\begin{equation} \label{eq_4.9}
\frac{d}{dt} \left(V^{(i+1)}(x) -W(x) \right) = 0. 
\end{equation}
This implies that $ V^{(i+1)}(x) -W(x) = c $ for $ \forall x \in \Omega $, where $ c $  is a real constant. For $ x=0 $, $ c = V^{(i+1)}(0) -W(0) = 0 $. Then,$ V^{(i+1)}(x) -W(x) = 0 $, i.e., $ W(x) = V^{(i+1)}(x)$   for $ \forall x \in \Omega $. From \eqref{eq_4.8}, we have that
\begin{flalign}
[\nu^{(i+1)}(x) - \upsilon(x)]^T R [\phi(\nu^{(i)}(x)) - u] = 0 \nonumber
\end{flalign}
for $ \forall u \in \mathcal{U}$, thus $ \nu^{(i+1)}(x) -\upsilon(x) = 0 $, i.e., $ \upsilon(x) = \nu^{(i+1)}(x)$   for $ \forall x \in \Omega $. This completes the proof. $ \square $

It follows from Theorem \ref{theorem_4.1} that the data-based policy iteration with equation \eqref{eq_4.3} is equivalent to the iteration of equations \eqref{eq_3.9}-\eqref{eq_3.11}, which is convergent as proved in \cite{abu2005nearly}. Thus, the convergence of the data-based policy iteration with equation \eqref{eq_4.3} can be guaranteed. 

\subsection{Data-based API based on actor-critic neural network structure} \label{Sec_4.2}
To solve equation \eqref{eq_4.3} for $ V^{(i+1)}(x) $  and $ \nu^{(i+1)}(x) $ based on data instead of system model, we develop an actor-critic NN-based approach, where critic and actor NNs are used to approximate cost function $ V^{(i)}(x) $  and policy $ \nu^{(i)}(x) $  respectively. From the well known high-order Weierstrass approximation theorem \cite{courant2004methods}, it follows that a continuous function can be accurately represented by an infinite-dimensional linearly independent basis function set. For real practical application, it is usually required to approximate the function in a compact set with a finite-dimensional function set.  We consider the critic and actor NNs for approximating the cost function and control policy on a compact set  $ \Omega $. Let $\varphi(x) \triangleq [\varphi_1(x)~...~\varphi_{L_V}(x)]^T$  be a vector of linearly independent activation functions for critic NN, where $\varphi_j(x): \Omega\mapsto \mathbb{R}, j=1,...,L_V, L_V$ is the number of critic NN hide layer neurons. Let $\psi^l(x) \triangleq [\psi_1^l(x)~...~\psi_{L_u}^l(x)]^T$ , be a vector of linearly independent activation functions of the $l$-th sub-actor NN for approximating policy $\nu_l, l=1,...,m$, where $\psi_k^l(x): \Omega\mapsto \mathbb{R}, k=1,...,L_u, L_u$ is the number of actor NN hide layer neurons. Then, the outputs of critic and the $l$-th sub-actor NNs are given by
\begin{flalign}
\widehat{V}^{(i)} (x) &= \sum_{l=1}^{L_V} \theta_{V,j}^{(i)} \varphi_j(x) = \varphi^T(x) \theta_V^{(i)}\label{eq_4.10}
\\
\widehat{\nu}_l^{(i)} (x) &= \sum_{k=1}^{L_u} \theta_{u_l,k}^{(i)} \psi_k^l(x) = (\psi^l(x))^T \theta_{u_l}^{(i)}\label{eq_4.11}
\end{flalign}
for $ \forall i=0,1,2,... $, where  $\theta_V^{(i)} \triangleq [\theta_{V,1}^{(i)}~...~\theta_{V,L_V}^{(i)}]^T$ and $\theta_{u_l}^{(i)} \triangleq [\theta_{u_l,1}^{(i)}~...~\theta_{u_l,L_u}^{(i)}]^T$  are weight vectors of critic and actor NNs respectively. Expression \eqref{eq_4.11} can be rewritten as a compact form
\begin{flalign}
\widehat {\nu}^{(i)} (x) &= \left[\widehat{\nu}_1^{(i)} (x)~...~\widehat{\nu}_m^{(i)} (x) \right]^T \nonumber \\
&= \left[(\psi^1(x))^T \theta_{u_1}^{(i)}~...~(\psi^m(x))^T \theta_{u_m}^{(i)}\right]^T. \label{eq_4.12}
\end{flalign}

Due to estimation errors of the critic and actor NNs \eqref{eq_4.10} and \eqref{eq_4.11}, the replacement of $ V^{(i+1)}$ and  $ \nu^{(i+1)}  $ in the iterative equation \eqref{eq_4.3} with $\widehat{V}^{(i+1)}$ and  $ \widehat{\nu}^{(i+1)} $ respectively, yields the following residual error:
\begin{flalign}\label{eq_4.13}
\sigma ^{(i)} (x(t), u(t)) \triangleq & 2\int_{t}^{t+\Delta t} [\widehat{\nu}^{(i+1)}(x(\tau))]^T R [\phi(\widehat{\nu}^{(i)}(x(\tau))) - u(x(\tau))] d \tau \nonumber\\
&+\widehat{V}^{(i+1)}(x(t)) - \widehat{V}^{(i+1)}(x(t+\Delta t)) 
- \int_{t}^{t+\Delta t} \left(Q(x(\tau)) + 2\int_{0}^{\phi(\widehat{\nu}^{(i)}(x(t)))} (\phi^{-1} (\mu))^T R d\mu \right)d\tau 
\end{flalign}
By using \eqref{eq_4.10} and \eqref{eq_4.12}, we have
\begin{flalign}\label{eq_4.14}
 \sigma ^{(i)} (x(t), u(t))  
=& \left[ \varphi(x(t)) - \varphi(x(t+\Delta t)) \right]^T \theta_{V}^{(i+1)}  
+ 2\sum_{l=1}^{m} r_l  
\left[ \int_{t}^{t+\Delta t} \phi\left(\left(\psi^l(x(t)) \right) ^T \theta_{u_l}^{(i)} \right) \left(\psi^l(x(\tau)) \right)^T d \tau \right] \theta_{u_l}^{(i+1)} \nonumber\\
& - 2\sum_{l=1}^{m} r_l \left[ \int_{t}^{t+\Delta t} u_l(x(\tau)) \left(\psi^l(x(\tau)) \right) ^T d \tau \right] \theta_{u_l}^{(i+1)}
-\int_{t}^{t+\Delta t} Q(x(\tau))d \tau \nonumber\\
& -2\sum_{l=1}^{m} r_l \int_{t}^{t+\Delta t} \left( \int_{0}^{\phi\left(\left(\psi^l(x(t)) \right) ^T \theta_{u_l}^{(i)} \right)} \phi^{-1} (\mu) d\mu \right) d\tau 
\end{flalign}
For notation simplicity, define
\begin{flalign}
\rho _ {\Delta \varphi} (x(t)) &\triangleq  \left[ \varphi(x(t)) - \varphi(x(t+\Delta t)) \right]^T \nonumber\\
\rho _ \psi ^{(i)l} (x(t)) &\triangleq  \int_{t}^{t+\Delta t} \phi\left(\left(\psi^l(x(t)) \right) ^T \theta_{u_l}^{(i)} \right) \left(\psi^l(x(\tau)) \right)^T d \tau \nonumber\\
\rho _ {u \psi}^l (x(t),u(t)) &\triangleq  \int_{t}^{t+\Delta t} u_l(x(\tau)) \left(\psi^l(x(\tau)) \right) ^T d \tau \label{eq_4.14a} \\
\rho _ Q (x(t)) &\triangleq  \int_{t}^{t+\Delta t} Q(x(\tau))d \tau\nonumber\\
\rho _ 1^{(i)l} (x(t)) &\triangleq  \int_{t}^{t+\Delta t} \left( \int_{0}^{\phi \left(\left(\psi^l(x(t)) \right) ^T \theta_{u_l}^{(i)} \right)} \phi^{-1} (\mu) d\mu \right) d\tau \nonumber
\end{flalign}
Then, equation \eqref{eq_4.14} is rewritten as
\begin{flalign}\label{eq_4.15a}
\sigma ^{(i)} (x(t), u(t)) 
=& \rho _ {\Delta \varphi} (x(t)) \theta_{V}^{(i+1)} 
+ 2\sum_{l=1}^{m} r_l \rho _ \psi ^{(i)l} (x(t)) \theta_{u_l}^{(i+1)} 
- 2\sum_{l=1}^{m} r_l \rho _ {u \psi}^l (x(t),u(t)) \theta_{u_l}^{(i+1)} \nonumber\\
& -\rho _ Q (x(t))  -2\sum_{l=1}^{m} r_l \rho _ 1^{(i),l} (x(t)) 
\end{flalign} 
To write equation \eqref{eq_4.15a} in a compact form, define
\begin{flalign}
\theta^{(i+1)} & \triangleq  \left[ \left(\theta_{V}^{(i+1)}\right)^T ~ \left(\theta_{u_1}^{(i+1)}\right)^T~...~\left(\theta_{u_m}^{(i+1)}\right)^T \right] \nonumber \\
\overline{\rho}_{u\psi}^{(i)l}(x(t),u(t)) & \triangleq r_l \left[ \rho _ \psi ^{(i)l} (x(t)) - \rho _ {u \psi}^l (x(t),u(t)) \right] \label{eq_4.15b}  \\
\overline{\rho}^{(i)}(x(t), u(t)) &\triangleq \left[ \rho ^T_ {\Delta \varphi} (x(t))~2\rho _ {u \psi}^{(i)1} (x(t),u(t))~...~2\rho _ {u \psi}^{(i)m} (x(t),u(t))\right] \nonumber\\
\overline{\rho}_1^{(i)}(x(t)) & \triangleq  \rho_Q (x(t))  +2\sum_{l=1}^{m} r_l \rho _ 1^{(i)l} (x(t)) \nonumber
\end{flalign} 
then, equation \eqref{eq_4.15a} is represented as
\begin{equation}\label{eq_4.15}
\sigma ^{(i)} (x(t), u(t))  = \overline{\rho}^{(i)}(x(t),u(t)) \theta^{(i+1)} - \overline{\rho}_1^{(i)}(x(t))
\end{equation}
Based on the method of weighted residuals \citep{finlayson1972method}, the unknown critic NN weight vector  $ \theta^{(i+1)} $ can be computed in such a way that residual error $ \sigma ^{(i)} (x, u) $  (for $ \forall t \geq 0 $) of \eqref{eq_4.15} is forced to be zero in some average sense. Thus, projecting the residual error $ \sigma ^{(i)} (x, u) $  onto $ d \sigma ^{(i)}/ d \theta^{(i+1)} $  and setting the result to zero on domain $ \mathcal {D} $  using the inner product, $ \langle \cdot , \cdot \rangle _{\mathcal {D} } $ , i.e., 
\begin{equation} \label{eq_4.16}
\left<d \sigma ^{(i)}/ d \theta^{(i+1)}, \sigma ^{(i)} (x, u) \right> _{\mathcal {D} } = 0.
\end{equation}
Then, the substitution of \eqref{eq_4.15} into \eqref{eq_4.16} yields, 
\begin{equation} 
\left< \overline{\rho}^{(i)}(x, u) , \overline{\rho}^{(i)}(x, u) \right> _{\mathcal {D} } \theta^{(i+1)} 
-  \left<\overline{\rho}^{(i)}(x, u),  \overline{\rho}_1^{(i)}(x) \right> _{\mathcal {D} } = 0 \nonumber \\
\end{equation}
and thus $ \theta^{(i+1)}  $ can be obtained with
\begin{equation} \label{eq_4.17}
\theta^{(i+1)} = \left< \overline{\rho}^{(i)}(x, u) , \overline{\rho}^{(i)}(x, u) \right> ^{-1} _{\mathcal {D} } 
\left<\overline{\rho}^{(i)}(x, u),  \overline{\rho}_1^{(i)}(x) \right> _{\mathcal {D} }.
\end{equation}

The computation of inner products $ \left< \overline{\rho}^{(i)}(x, u) , \overline{\rho}^{(i)}(x, u) \right> _{\mathcal {D} } $ and $ \left<\overline{\rho}^{(i)}(x, u),  \overline{\rho}_1^{(i)}(x) \right> _\mathcal {D} $ involve many numerical integrals on domain $ \mathcal {D} $, which are computationally expensive. Thus, the Monte-Carlo integration method \citep{peter1978new} is introduced, which is especially competitive on multi-dimensional domain. We now illustrate the Monte-Carlo integration for computing $ \left< \overline{\rho}^{(i)}(x, u) , \overline{\rho}^{(i)}(x, u) \right> _{\mathcal {D} } $. Let  $ I_\mathcal {D} \triangleq \int_\mathcal {D} d(x, u)  $, and $ \mathcal {S}_M \triangleq \lbrace (x_k, u_k) \vert (x_k, u_k) \in \mathcal {D}, k = 1,2,...,M \rbrace  $   be the set that sampled on domain  $ \mathcal {D} $, where $ M $ is size of sample set $ \mathcal {S}_M $. Then, $ \left< \overline{\rho}^{(i)}(x, u) , \overline{\rho}^{(i)}(x, u) \right> _{\mathcal {D} } $  is approximately computed with
\begin{flalign} \label{eq_4.18}
\left< \overline{\rho}^{(i)}(x, u) , \overline{\rho}^{(i)}(x, u) \right> _\mathcal {D}
& = \int _\mathcal {D} \left( \overline{\rho}^{(i)}(x, u) \right)^T \overline{\rho}^{(i)}(x, u) d(x,u)\nonumber \\
& = \frac{I_\mathcal {D}}{M} \sum _{k=1}^M \left( \overline{\rho}^{(i)}(x_k, u_k) \right)^T \overline{\rho}^{(i)}(x_k, u_k) \nonumber \\
& = \frac{I_\mathcal {D}}{M} \left( Z^{(i)} \right)^T Z^{(i)}
\end{flalign} 
where $ Z^{(i)} \triangleq \left[ \left( \overline{\rho}^{(i)}(x_1, u_1) \right)^T~...~\left( \overline{\rho}^{(i)}(x_M, u_M) \right)^T \right]^T $. Similarly, 
\begin{flalign} \label{eq_4.19}
\left<\overline{\rho}^{(i)}(x, u),  \overline{\rho}_1^{(i)}(x) \right> _\mathcal {D}
& = \frac{I_\mathcal {D}}{M} \sum _{k=1}^M \left( \overline{\rho}^{(i)}(x_k, u_k) \right)^T \overline{\rho}_1^{(i)}(x_k)\nonumber \\
& = \frac{I_\mathcal {D}}{M} \left( Z^{(i)} \right)^T \eta^{(i)}
\end{flalign}
where $ \eta^{(i)} \triangleq \left[ \overline{\rho}_1^{(i)}(x_1)~...~\overline{\rho}_1^{(i)}(x_M) \right]^T $. Then, the substitution of \eqref{eq_4.18} and \eqref{eq_4.19} into \eqref{eq_4.17} yields,
\begin{equation} \label{eq_4.20}
\theta^{(i+1)} = \left( Z^{(i)} \right)^T Z^{(i)} \left( Z^{(i)} \right)^T \eta^{(i)}.
\end{equation}

Obviously, expression \eqref{eq_4.20} is a least-square scheme. Here, the sample set $ \mathcal {S}_M $  is collected from neighborhood of system state trajectories under a control policy with exploratory noise. Let  $ t_k = (k-1) \Delta t, (k=1,...,M) $,  $ x(t_k) $ and $ u(t_k) $  are the system state and control action at time instant  $ t_k $. Select the sample set $ \mathcal {S}_M = \lbrace (x_k, u_k) \vert (x_k, u_k) = (x(t_k), u(t_k)), k = 1,2,...,M \rbrace  $  for computing $ Z^{(i)} $  and $ \eta^{(i)} $  in \eqref{eq_4.20}. After $ Z^{(i)} $  and $ \eta^{(i)} $ are computed,  $ \theta^{(i+1)} $ can be obtained accordingly. 

\noindent \textbf{Remark 4.} Note that the least-square method \eqref{eq_4.20} requires the inverse of matrix  $ (Z^{(i)} )^T Z^{(i)} $, i.e., $ Z^{(i)} $  should be full column rank, which can be realized from two aspects in the practical implementation. 1) It is noted that $ \theta^{(i+1)} $  has $ L_V + mL_u$  unknown parameters. This means that, in order to solve for $ \theta^{(i+1)} $  with least-square scheme \eqref{eq_4.20} , it is practical to increase the size of sample set $ \mathcal {S}_M $  such that  $ M \gg L_V + mL_u$. 2) Choose the persistent exciting input signal $ u $ that contains enough frequencies, which is similar with the issue ``exploration" of RL in machine community. $ \square $

\subsection{Implementation of the data-based API algorithm} \label{Sec_4.3}
In the above subsection \ref{Sec_4.2}, the developed least-square scheme \eqref{eq_4.20} is designed only for solving one iterative equation \eqref{eq_4.3}. Now, we present a complete data-based API algorithm procedure for  constrained optimal control design as follows:

\begin{algorithm} \label{algorithm_4.1}
Data-based API algorithm for constrained optimal control design.
\begin{itemize}
\item Step 1: Select an initial actor NN weight vector $ \theta_{u_l}^{(0)} (l=1,...,m) $ such that $ \phi( \widehat {\nu}^{(0)}) \in \mathfrak{U}(\Omega) $. Use the input signal $ u = \phi(\nu), \nu = \widehat {\nu}^{(0)} +e_u $  to the system \eqref{eq_2.1} for closed-loop simulation, where $ e_u $  is the exploratory noise. Measure system state and input signal online for sample set $ \mathcal {S}_M $, and compute $ \rho _ {\Delta \varphi} (x_k), \rho _ Q (x_k), \rho _ {u \psi}^{l} (x_k,u_k), k=1,...,M $;
\item Step 2: Set initial critic NN weight  $ \theta_{V}^{(0)}=0$. Let $ i = 0 $;
\item Step 3: Compute $ Z^{(i)} $ and $\eta^{(i)} $, and update $\theta^{(i+1)} $ with \eqref{eq_4.20};
\item Step 4: Let $i = i+1 $.  If $ \Vert \theta^{(i)} - \theta^{(i-1)} \Vert \leq \xi $ ($ \xi $ is a small positive number), stop iteration and $\theta^{(i)} $ is employed to obtain the final control policy $ \phi(\widehat{\nu}^{(i)}) $, else go back to Step 3 and continue. $\square $
\end{itemize}
\end{algorithm}

\noindent \textbf{Remark 5.} It is found that the data-based API algorithm uses online state and input information of the closed-loop system instead of dynamic model, for learning the optimal control policy \eqref{eq_3.5} and the solution of HJB equation \eqref{eq_3.4}. The procedure of API algorithm can be divided into an online and an offline part. 1) Step 1 is online part for data processing. By collecting system state and input signal  for sample set $ \mathcal {S}_M $, compute $ \rho _ {\Delta \varphi} (x_k), \rho _ Q (x_k) $ and $ \rho _ {u \psi}^{l} (x_k,u_k) $, and then prepare for iteration. In fact, the information of the system dynamics is embedded in the data measured online, and thus explicit system identification is avoided. 2) Steps 2-4 is the offline part for iterative learning the optimal control policy and the solution of HJB equation. After the iteration is convergent, the resulting actor NN weight is applied to obtain the optimal control policy for real control.$ \square $

\section{Data-based approximate policy iteration for unconstrained optimal control} \label{Sec_5}
In this section, the developed data-based API algorithm is simplified for solving the unconstrained optimal control problem of system \eqref{eq_2.1}. The derivation of the algorithm is similar with that for constrained optimal control design in Section \ref{Sec_4}, thus the procedure is presented briefly.  Letting $ V^{(i+1)}(x) $ be the solution of the LFE \eqref{eq_3.14} with \eqref{eq_3.15}, take derivative of $ V^{(i+1)}(x) $  with respect to time along the state of system \eqref{eq_4.1}
\begin{flalign}\label{eq_5.2}
\frac{dV^{(i+1)}(x)}{dt} &= [\nabla V^{(i+1)}]^T (f + gu^{(i)} ) + [\nabla V^{(i+1)}]^T g[u - u^{(i)}] \nonumber \\
&=  -Q(x) - \Vert u^{(i)} \Vert _R^2 + 2[u^{(i+1)}]^T R [u^{(i)} - u].
\end{flalign}
Integrating both sides of \eqref{eq_5.2} on the interval $ [t, t+\Delta t] $  and rearranging terms yields, 
\begin{eqnarray}\label{eq_5.3}
V^{(i+1)}(x(t)) - V^{(i+1)}(x(t+\Delta t))
+  2\int_{t}^{t+\Delta t}  [u^{(i+1)}(x(\tau))]^T R [ u^{(i)}(x(\tau)) - u(\tau)] d \tau \nonumber \\
= \int_{t}^{t+\Delta t} [Q(x(\tau)) + \Vert u^{(i)} (x(\tau)) \Vert _R^2] d \tau.
\end{eqnarray}
The convergence of the data-based policy iteration with \eqref{eq_5.3} is summarized in Theorem \ref{theorem_5.1}.

\begin{theorem}\label{theorem_5.1}
Let $ V^{(i+1)}(x) \in C^1(\Omega), V^{(i+1)}(x) \geq 0,  V^{(i+1)}(0) = 0 $ and $ u^{(i+1)}(x) \in \mathfrak{U}(\Omega) $. $ (V^{(i+1)}(x), u^{(i+1)}(x)) $ is the solution of equation \eqref{eq_5.3} iff $($ if and only if $ ) $ it is the solution of the LFE \eqref{eq_3.14} and \eqref{eq_3.15}, i.e., equation \eqref{eq_5.3} is equivalent to the LFE \eqref{eq_3.14} with \eqref{eq_3.15}.$ \square $
\end{theorem} 
Theorem \ref{theorem_5.1} can be easily proved similar with the Proof of Theorem \ref{theorem_4.1}, thus it is omitted for brevity. With the same critic and actor NN structures \eqref{eq_4.10} and \eqref{eq_4.11} for estimating $ V^{(i)}(x) $ and $ u^{(i)}(x) $, the replacement of $ V^{(i+1)}$ and  $ u^{(i+1)}  $ in the iterative equation \eqref{eq_5.3} with $ \widehat{V}^{(i+1)}$ and  $ \widehat{u}^{(i+1)}  $ respectively, yields the following residual error:
\begin{flalign}\label{eq_5.14}
\sigma ^{(i)} (x(t), u(t)) 
=&[\varphi(x(t)) - \varphi(x(t+\Delta t))]^T \theta_V^{(i+1)} 
+ 2 \sum_{l=1}^m r_{l} \int_{t}^{t+\Delta t} [(\psi^{l}(x(\tau)))^T \theta_{u_{l}}^{(i)} - u_{l}(\tau)] (\psi^{l}(x(\tau)))^T \theta_{u_{l}}^{(i+1)} d \tau \nonumber \\
& -\int_{t}^{t+\Delta t} Q(x(\tau)) d \tau - \sum_{l=1}^m r_{l} \int_{t}^{t+\Delta t} (\theta_{u_{l}}^{(i)})^T \psi^{l}(x(\tau))  (\psi^{l}(x(\tau)))^T \theta_{u_{l}}^{(i)}  d \tau.
\end{flalign}
With the notations $ \rho _ {\Delta \varphi} (x(t)), \rho _ {u \psi}^l (x(t),u(t)), \rho _ Q (x(t)) $ defined in \eqref{eq_4.14a}, and let $ \rho _ \psi ^l (x(t)) $ be
\begin{equation}
\rho _ \psi ^l (x(t)) \triangleq \int_{t}^{t+\Delta t} \psi^{l}(x(\tau))  (\psi^{l}(x(\tau)))^T  d \tau
\nonumber\\
\end{equation}
the expression \eqref{eq_5.14} is rewritten as
\begin{flalign}\label{eq_5.15}
\sigma ^{(i)} (x(t), u(t)) 
=& \rho ^T_ {\Delta \varphi} (x(t)) \theta_V^{(i+1)} 
+ 2 \sum_{l=1}^m  r_{l} \left[ (\theta_{u_{l}}^{(i)})^T \rho _ \psi^{l} (x(t)) - \rho _ {u\psi}^{l} (x(t),u(t)) \right] \theta_{u_{l}}^{(i+1)}  \nonumber \\
& -\rho _ Q (x(t)) - \sum_{l=1}^m  r_{l} (\theta_{u_{l}}^{(i)})^T \rho _ \psi^{l} (x(t)) \theta_{u_{l}}^{(i)} \nonumber \\
 = & \overline{\rho}^{(i)}(x(t), u(t)) \theta^{(i+1)} -\overline{\rho}_1^{(i)}(x(t))
\end{flalign}
where the notations $ \theta^{(i+1)}, \overline{\rho}^{(i)}(x(t), u(t)) $ are defined in \eqref{eq_4.15b}, $ \overline{\rho}_1^{(i)}(x(t)) $ and $ \rho _ {u \psi}^{(i)l} (x(t),u(t)) $ are given by
\begin{flalign}
\overline{\rho}_1^{(i)}(x(t)) &= \rho _ Q (x(t)) + \sum_{l=1}^m r_{l}  (\theta_{u_{l}}^{(i)})^T \rho _ \psi^{l} (x(t)) \theta_{u_{l}}^{(i)} \nonumber \\
\rho _ {u \psi}^{(i)l} (x(t),u(t)) & = r_{l} \left[ (\theta_{u_{l}}^{(i)})^T \rho _ \psi^{l} (x(t)) - \rho _ {u\psi}^{l} (x(t),u(t)) \right] .  \nonumber
\end{flalign}
Note that the expression \eqref{eq_5.15} is the same as \eqref{eq_4.15}, thus with the method of weighted residuals described in Subsection \ref{Sec_4.2}, the least-square scheme \eqref{eq_4.20} can also be obtained for computing unknown parameter vector $ \theta^{(i+1)} $. 

\begin{algorithm} \label{algorithm_5.1}
Data-based API algorithm for unconstrained optimal control design.
\begin{itemize}
\item Step 1: Select an initial actor NN weight vector $ \theta_{u_l}^{(0)} (l=1,...,m) $ such that $ \widehat {u}^{(0)} \in \mathfrak{U}(\Omega) $. Use the input signal $ u = \widehat {u}^{(0)} +e_u $  to the system \eqref{eq_2.1} for closed-loop simulation, where $ e_u $  is the exploratory noise. Measure system state and input signal online for sample set $ \mathcal {S}_M $, and compute $ \rho _ {\Delta \varphi} (x_k), \rho _ Q (x_k), \rho _ \psi^{l} (x_k), \rho _ {u \psi}^{l} (x_k,u_k), k=1,...,M $;
\item Step 2: Set initial critic NN weight  $ \theta_{V}^{(0)}=0$. Let $ i = 0 $;
\item Step 3: Compute $ Z^{(i)} $ and $\eta^{(i)} $, and update $\theta^{(i+1)} $ with \eqref{eq_4.20};
\item Step 4: Let $i = i+1 $.  If $ \Vert \theta^{(i)} - \theta^{(i-1)} \Vert \leq \xi $ ($ \xi $ is a small positive number), stop iteration and $\theta^{(i)} $ is employed to obtain the final control policy $ \widehat {u}^{(i)} $, else go back to Step 3 and continue. $\square $
\end{itemize}
\end{algorithm}

Next, we discuss the developed data-based API algorithm for special unconstrained linear systems. Consider the linear version of system \eqref{eq_2.1}:
\begin{equation}\label{eq_5.21}
\dot{x}(t) = Ax(t) + Bu(t), x(0) = x_0 
\end{equation}
and linear quadratic cost function:
\begin{equation} \label{eq_5.22}
V(x_0) = \int_{0}^{\infty} {\left( \Vert x(t) \Vert _Q^2 + \Vert u(t) \Vert _R^2 \right)} dt
\end{equation}
where  $ A,B $ are matrices of appropriate dimension, and$ Q>0 $. From the linear quadratic regulator theory \cite{lewis2013optimal, anderson2007optimal}, the solution of HJB equation \eqref{eq_3.2} is  $ V^*(x) = x^TPx $, where $ P>0 $  is the solution of the algebraic Riccati equation:
\begin{equation} \label{eq_5.23}
A^TP + PA + Q  - PBR^{-1}B^TP =0.
\end{equation}
Let  $ V^{(i)}(x) = x^TP^{(i)}x $, then expression \eqref{eq_3.15} is given by $ u^{(i)}(x) = -K^{(i)}x $  with  $ K^{(i)} =  R^{-1}B^TP^{(i)}$. Thus, the iterative equation \eqref{eq_5.3} is rewritten as
\begin{eqnarray}\label{eq_5.24}
x(t)^TP^{(i+1)}x(t) - x(t+\Delta t)^TP^{(i+1)}x(t+\Delta t)
+  2\int_{t}^{t+\Delta t}  [ K^{(i)}x(\tau) + u(\tau)]^T R K^{(i+1)}x(\tau) d \tau \nonumber \\
= \int_{t}^{t+\Delta t} x^T(\tau) [Q + (K^{(i)})^T R K^{(i)}] x(\tau) d \tau.
\end{eqnarray}
It is observed that the iterative equation \eqref{eq_5.24} is the same as the iterative equation (10) in reference \cite{Jiang2012computational}. This means that for the unconstrained optimal control problem of linear systems, the developed data-based API algorithm results in the method in reference \cite{Jiang2012computational}.

\begin{figure}[htbp]
\centering\includegraphics[width=2.5in]{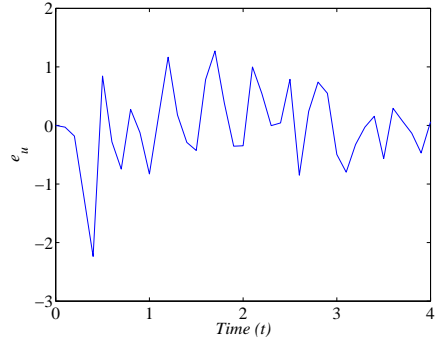}
\caption{For case1, the exploratory noise $ e_u $.}\label{fig1}
\end{figure}

\begin{figure}
	\begin{minipage}[t]{0.5 \linewidth}
	\setcaptionwidth{2.5in}
		\centering	\includegraphics[width=2.5in]{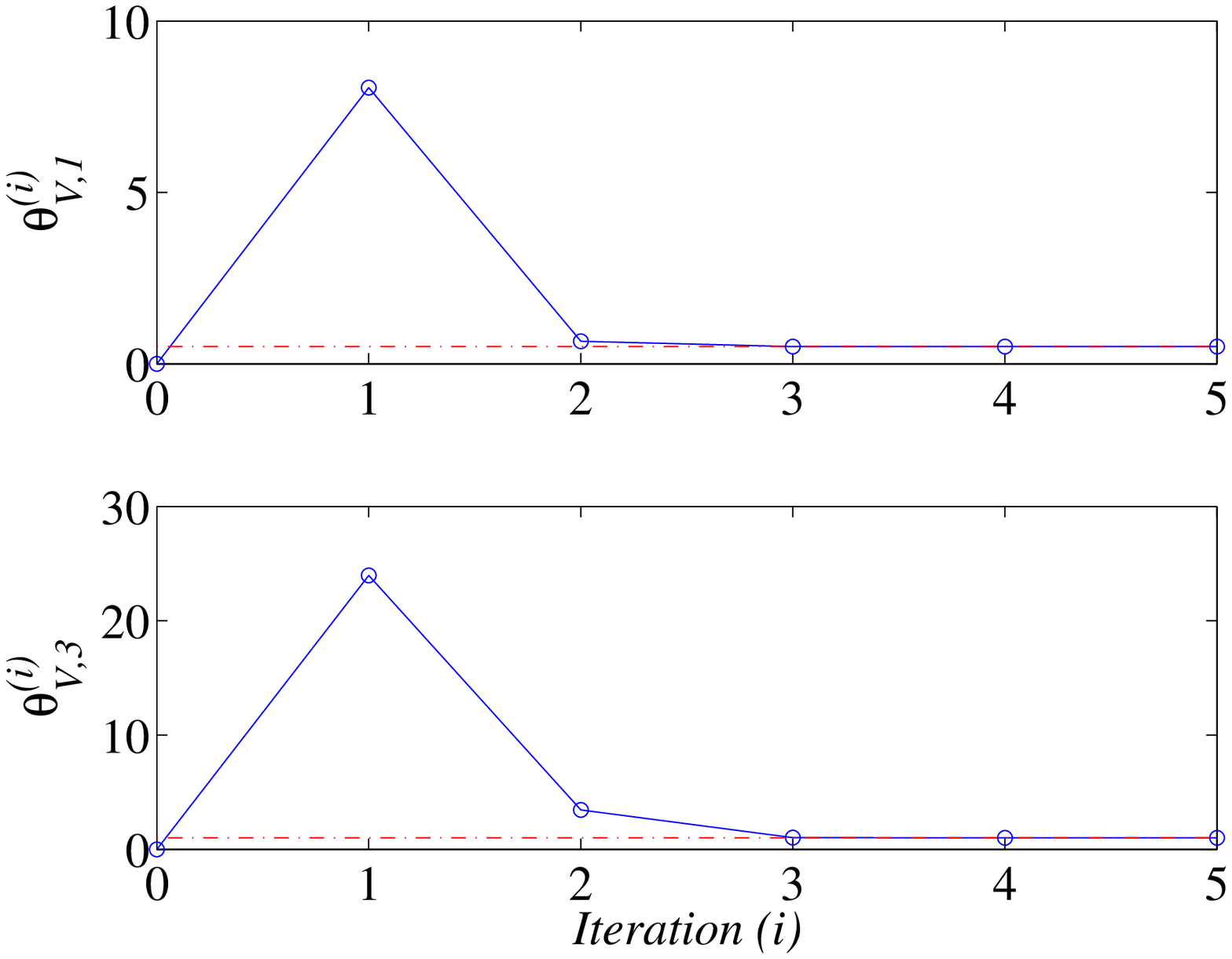}
	\caption{For case 1, two representative critic NN weights $ \theta_{V,1}^{(i)} $ and $ \theta_{V,3}^{(i)} $.} 
		\label{fig2}
	\end{minipage}%
	\begin{minipage}[t]{0.5\linewidth}
	\setcaptionwidth{2.5in}
		\centering	\includegraphics[width=2.5in]{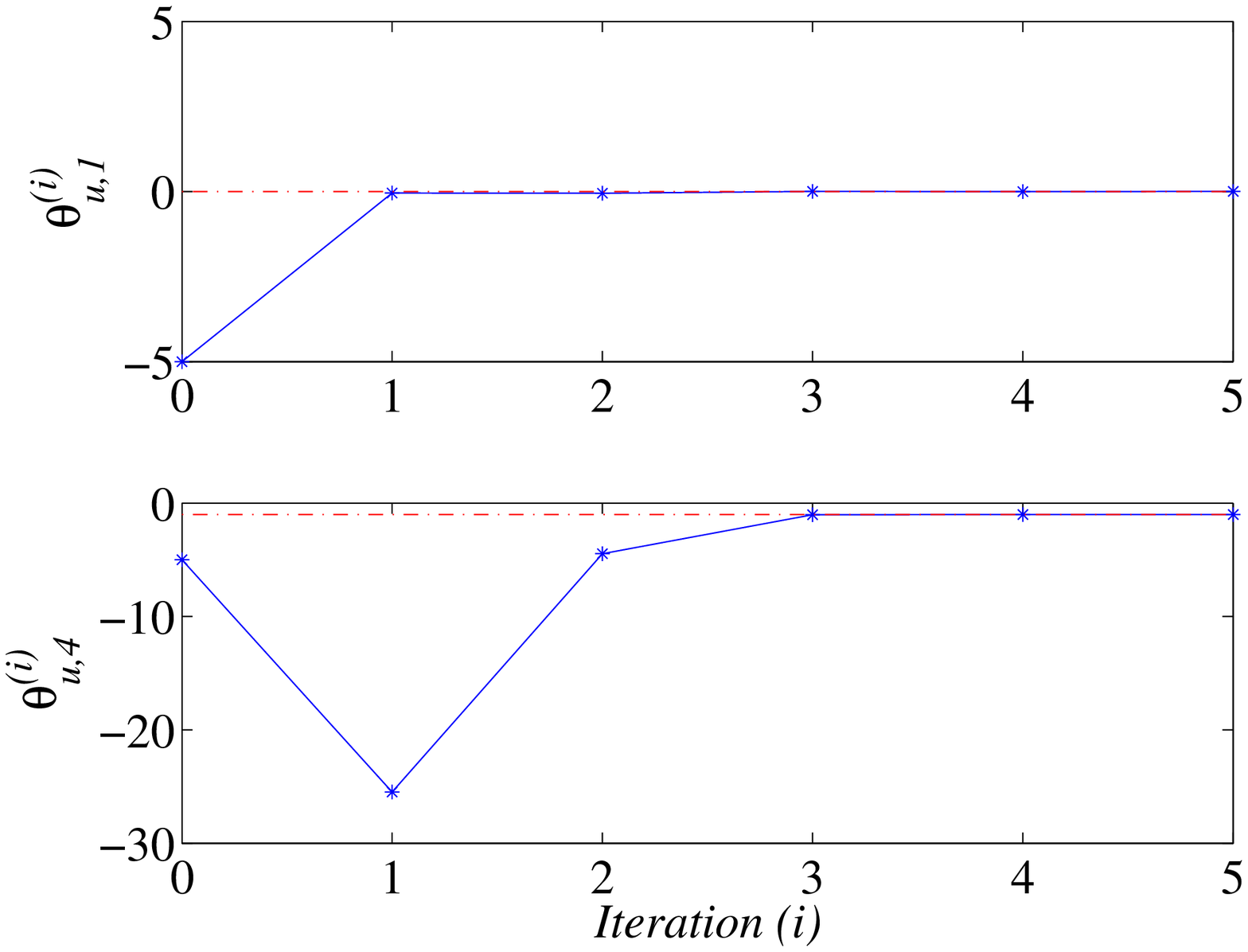}
		\caption{For case 1, two representative actor NN weights $ \theta_{u,1}^{(i)} $ and $ \theta_{u,4}^{(i)} $.}	
		\label{fig3}
	\end{minipage}
\end{figure}

\begin{figure}
	\begin{minipage}[t]{0.5 \linewidth}
	\setcaptionwidth{2.5in}
		\centering	\includegraphics[width=2.5in]{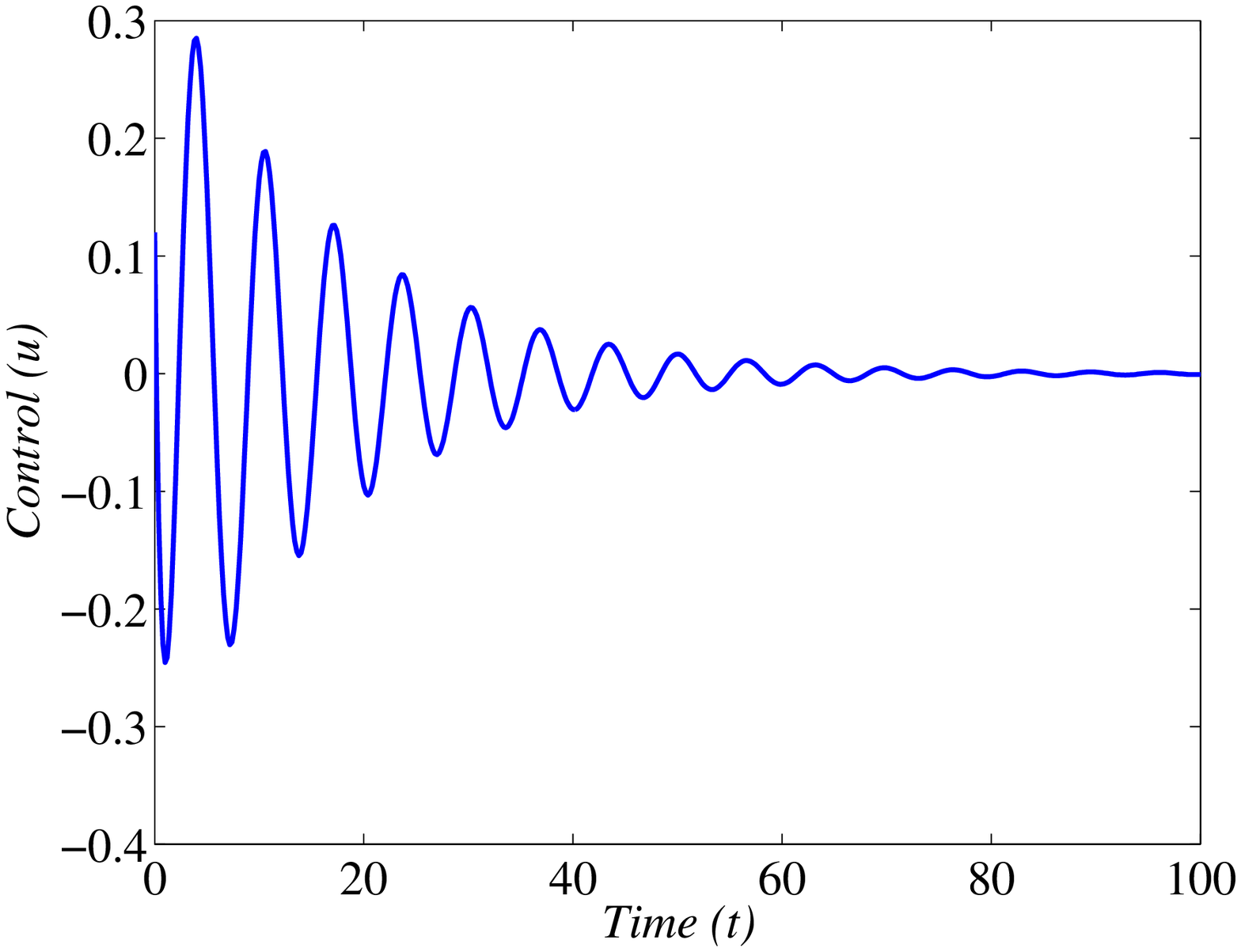}
	\caption{For case 2, trajectory of the initial control policy $ \widehat {u}^{(0)} $.}
		\label{fig4}
	\end{minipage}%
	\begin{minipage}[t]{0.5\linewidth}
	\setcaptionwidth{2.5in}
		\centering	\includegraphics[width=2.5in]{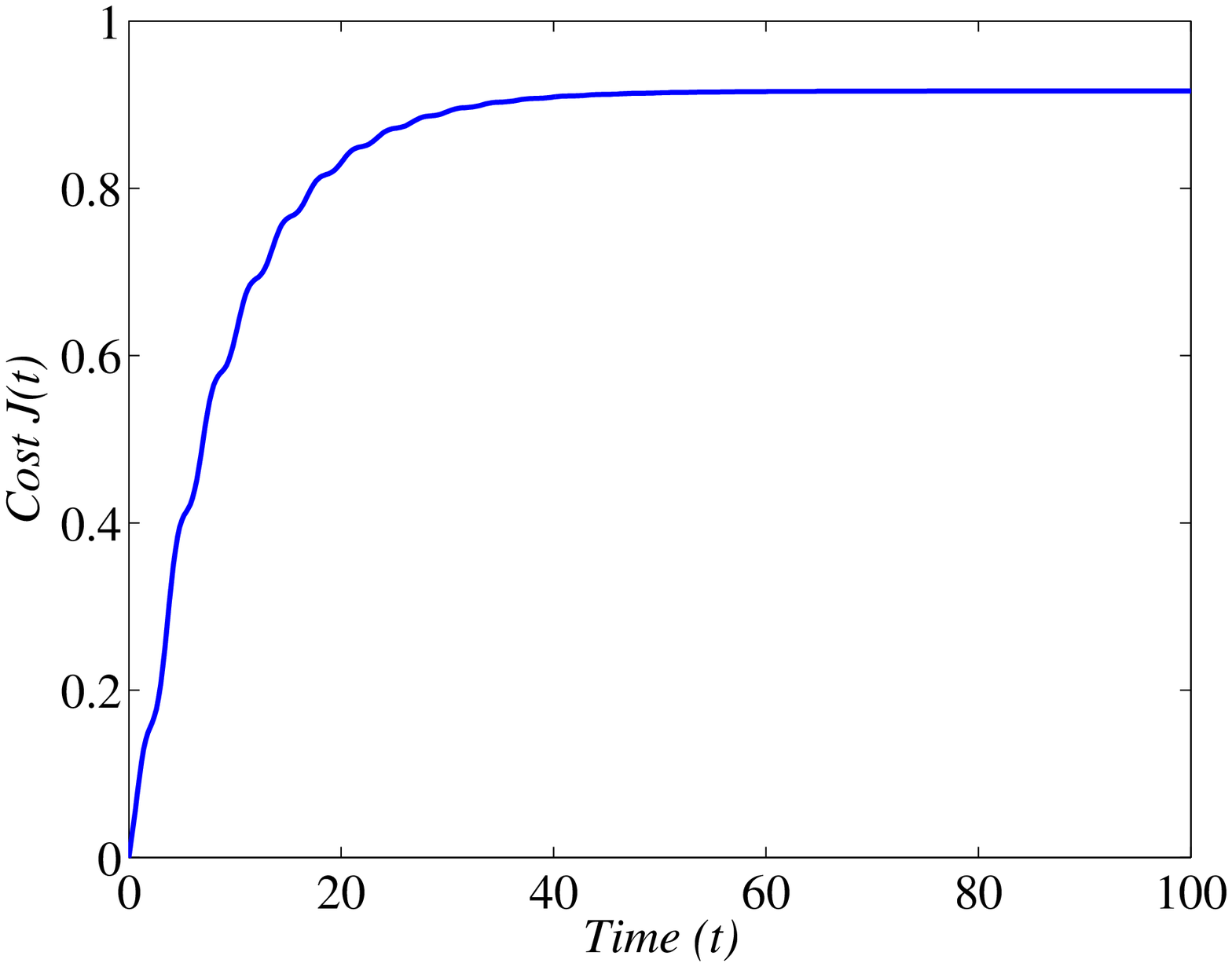}
		\caption{For case 2, trajectory of cost $J(t)$ with the initial control policy $ \widehat {u}^{(0)} $.}
		\label{fig5}
	\end{minipage}
\end{figure}

\section{Simulation studies} \label{Sec_6}
In this section, we first test the effectiveness of the developed data-based API algorithm on a simple unconstrained nonlinear numerical system, and further apply it to the complex RTAC nonlinear benchmark problem for both unconstrained and constrained optimal control design. 
\subsection{Case 1: Effectiveness test on a simple nonlinear numerical system} \label{Sec_6.1}
This numerical example is constructed by using the converse HJB approach \cite{nevistic1996optimality}. The system model is given as follows: 
\begin{eqnarray}  \label{eq_6.1}
\dot{x} = \left[ \begin{array} {*{3}{>{\displaystyle}c}}
 		-x_1 + x_2\\
 		-0.5(x_1 + x_2) + 0.5x_1^2x_2
	\end{array} \right] +
\left[ \begin{array} {*{3}{>{\displaystyle}c}}
 		0 \\
 		x_1
	\end{array} \right] u,x_0 =
\left[ \begin{array} {*{3}{>{\displaystyle}c}}
 		0.1 \\
 		0.1
	\end{array} \right]
\end{eqnarray}
With the choice of $ Q(x)=x^Tx $  and $ W(u) = u^2 $  for the cost function \eqref{eq_2.2}. From the converse HJB approach \cite{nevistic1996optimality}, the solution of the associated HJB equation \eqref{eq_3.12} is $ V^*(x) = 0.5x_1^2 + x_2^2 $, and thus  $ u^*(x) = -x_1x_2 $.

\begin{figure}[htbp]
\centering\includegraphics[width=5in]{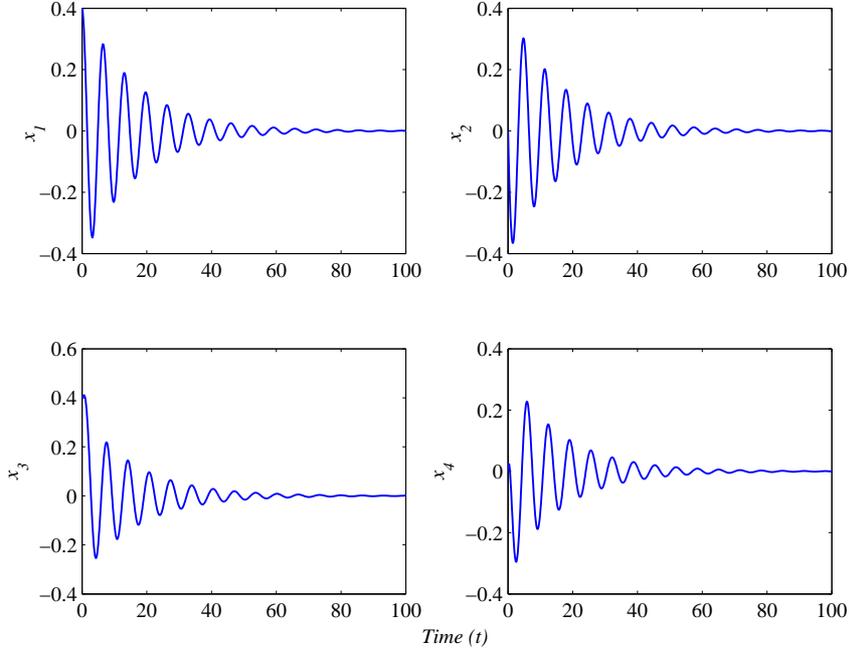}
\caption{For case 2, system state trajectories with the initial control policy $ \widehat {u}^{(0)} $.}\label{fig6}
\end{figure}

To solve the unconstrained optimal control problem with the data-based API algorithm (Algorithm \ref{algorithm_5.1}), select the critic NN activation function vector as $ \varphi (x) = [x_1^2~x_1x_2~x_2^2 ]^T $  with the size of  $ L_V = 3 $, actor NN activation function vector as  $ \psi (x) = [x_1~x_2~ x_1^2~x_1x_2~x_2^2 ]^T $ with the size of  $ L_u = 5 $, and the initial actor NN weight vector as $ \theta_{u}^{(0)} = [-5~-5~ -5~-5~-5]^T$. Since $ V^*(x) = 0.5x_1^2 + x_2^2 $  and  $ u^*(x) = -x_1x_2 $, the optimal critic and actor NN weight vectors are $ \theta_{V}^{*} = [0.5~0~ 1]^T$  and  $ \theta_{u}^{*} = [0~0~ 0~-1~0]^T$, respectively. To generate sample set   $ \mathcal {S}_M $ and compute $ \rho _ {\Delta \varphi} (x_k), \rho _ Q (x_k), \rho _ \psi^{l} (x_k), \rho _ {u \psi}^{l} (x_k,u_k)$, let sample size  $ M = 41 $ and time interval  $ \Delta t = 0.1s $. 
Then, we conducted closed-loop simulation on system \eqref{eq_6.1} with input signal $ u = \widehat {u}^{(0)} +e_u $, where $ e_u $ is exploratory noise generated by
\begin{equation} \label{eq_6.2}
e_u(t) = 0.05 \sum_{k=1} ^{100} \sin r_k t
\end{equation} 
with $ r_k \in [-100,100], (k = 1,...,100) $ be random parameters. Figure \ref{fig1} gives the noise signal $ e_u $.  
After the online procedure (i.e., Step 1) is completed, offline iteration (i.e., Steps 2-4) is used to learn the optimal control policy. Setting the value of convergence criterion  $ \xi = 10^{-5} $, it is found that the critic and actor NN weight vectors converge respectively to $ \theta_{V}^{*}  $  and  $ \theta_{u}^{*}$, at the $5^{th}$ iteration. Figure \ref{fig2} shows two representative critic NN weights   $ \theta_{V,1}^{(i)} $ and $ \theta_{V,3}^{(i)} $, and Figure \ref{fig3} demonstrates two representative actor NN weights $ \theta_{u,1}^{(i)} $ and $ \theta_{u,4}^{(i)} $, wherein the dashed lines are optimal values. By using the convergent actor NN weight vector  $ \theta_{u}^{(5)} $, closed-loop simulation is conducted with final control policy $ \widehat {u}^{(5)} $, and the real cost \eqref{eq_2.2} is 0.0150. Thus, the simulation on this simple nonlinear system demonstrates the effectiveness of the developed data-based API algorithm. 

\subsection{Case 2: Application to the unconstrained RTAC nonlinear benchmark problem} \label{Sec_6.2}
The rotational/translational actuator (RTAC) nonlinear benchmark problem has been used to test the abilities of control methods \citep{abu2008neurodynamic}. The dynamics of this nonlinear plant poses challenges as the rotational and translation motions are coupled. The RTAC system is given as follows:

\begin{figure}[htbp]
\centering\includegraphics[width=2.5in]{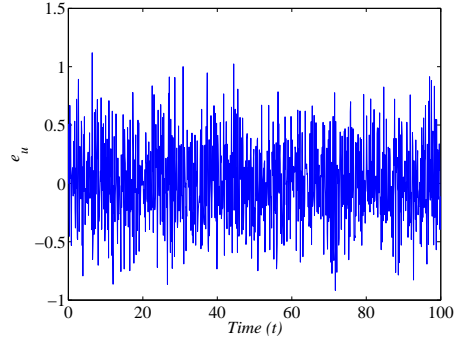}
\caption{For case 2, the exploratory noise.}\label{fig7}
\end{figure}

\begin{eqnarray}  \label{eq_6.3}
\dot{x} = \left[ \begin{array} {*{3}{>{\displaystyle}c}}
 		x_2\\
 		\frac{-x_1 + \zeta x_4^2 \sin x_3}{1 - \zeta ^2 \cos^2 x_3} \\
 		x_4\\
 		\frac{\zeta \cos x_3 (x_1 - \zeta x_4^2 \sin x_3)}{1 - \zeta ^2 \cos^2 x_3}
	\end{array} \right] +
\left[ \begin{array} {*{3}{>{\displaystyle}c}}
 		0 \\
 		 \frac{- \zeta \cos x_3}{1 - \zeta ^2 \cos^2 x_3} \\
 		0\\
 		\frac{1}{1 - \zeta ^2 \cos^2 x_3}
	\end{array} \right] u,  x_0=
\left[ \begin{array} {*{3}{>{\displaystyle}c}}
 		0.4 \\
 		0.0\\
 		0.4\\
 		0.0
	\end{array} \right]
\end{eqnarray}
where  $ \zeta =0.2 $. For the cost function \eqref{eq_2.2}, let $ W(u)=u^2 $  and $ Q(x)=x^T S x $ with $ S = diag(0.5~0.05~0.05~0.05)$.

\begin{figure}[htbp]
\centering\includegraphics[width=5.0in]{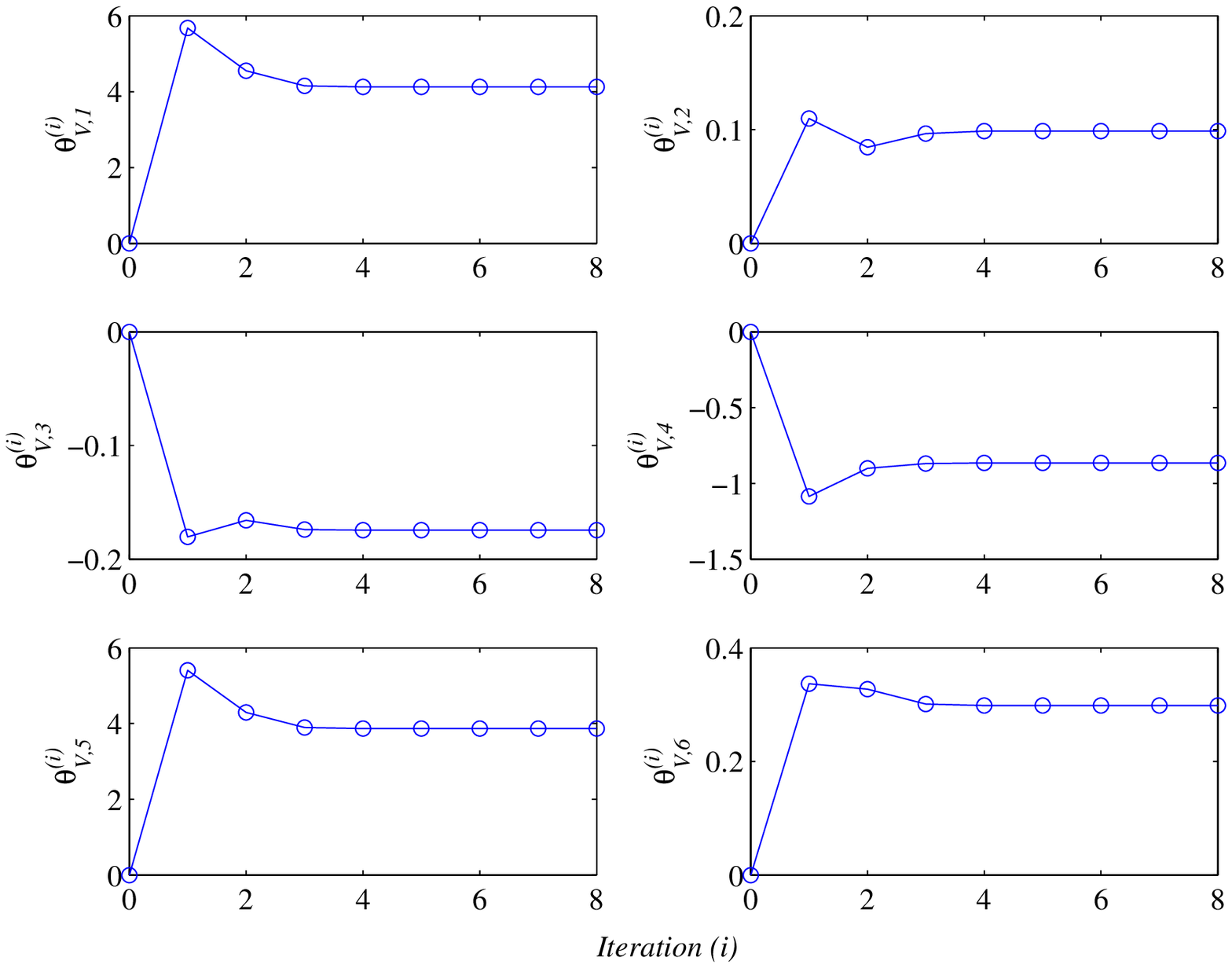}
\caption{For case 2, the first six representative critic NN weights $ \theta_{V,1}^{(i)}-\theta_{V,6}^{(i)} $ at each iteration. }\label{fig8}
\end{figure}

\begin{figure}[htbp]
\centering\includegraphics[width=5.0in]{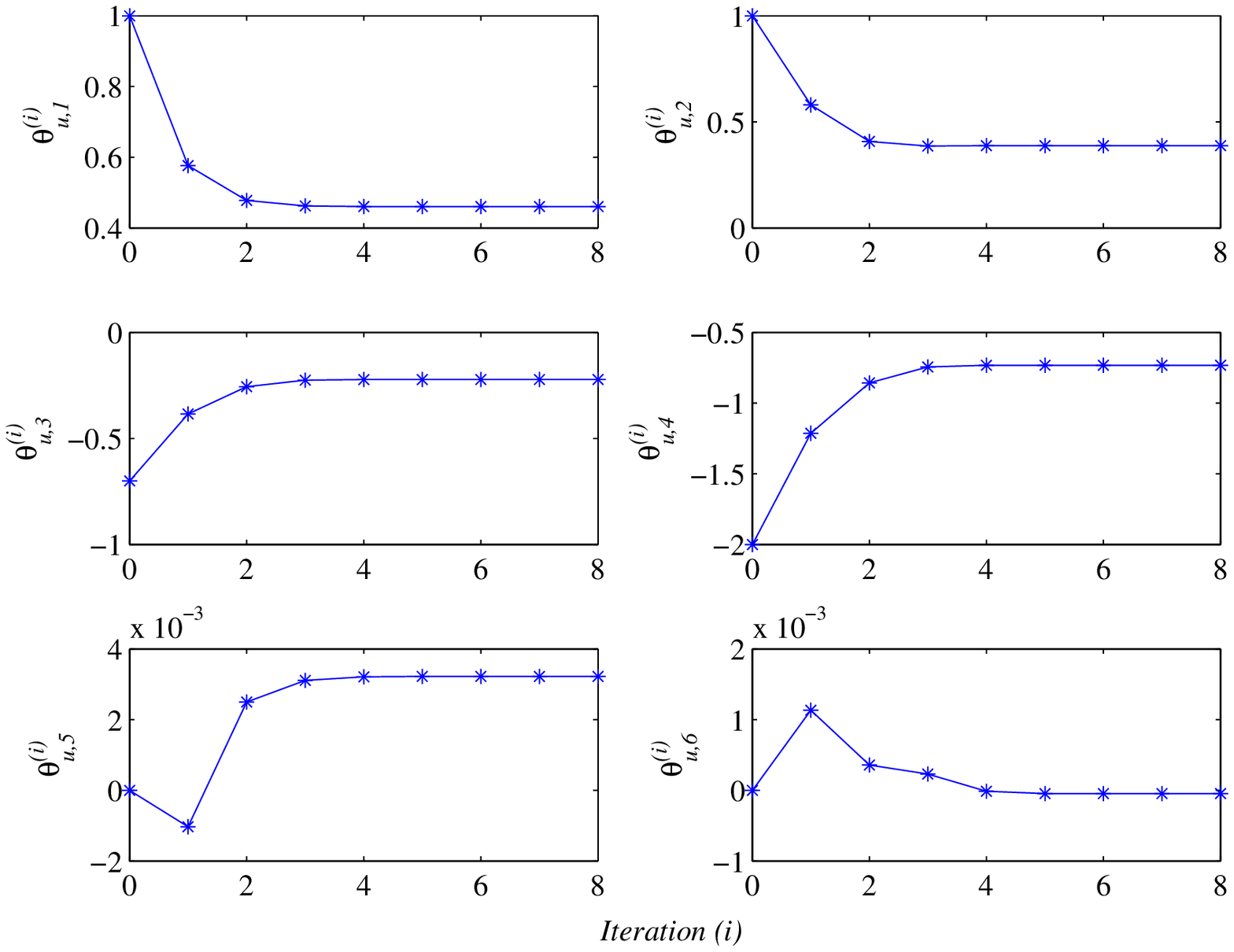}
\caption{For case 2, the first six representative actor NN weights $ \theta_{u,1}^{(i)}-\theta_{u,6}^{(i)} $ at each iteration.} \label{fig9}
\end{figure}

To learn the unconstrained optimal control policy with the data-based API algorithm (Algorithm \ref{algorithm_5.1}), select the critic NN activation function vector as
\begin{equation} \label{eq_6.4}
 \begin{array}{cccccccccc}
 \varphi(x) = [
 x_1^2 & x_1x_2 & x_1x_3 
 & x_1x_4 & x_2^2 & x_2x_3 
 & x_2x_4 & x_3^2 & x_3x_4
 & x_4^2 \\ x_1^3x_2 & x_1^3x_3 
 & x_1^3x_4 & x_1^2x_2^2 & x_1^2x_2x_3 
 & x_1^2x_2x_4 & x_1^2x_3^2 & x_1^2x_3x_4
 & x_1^2x_4^2 & x_1x_2^3 \\ x_1x_2^2x_3 
 & x_1x_2^2x_4 & x_1x_2x_3^2 & x_1x_2x_3x_4 
 & x_1x_2^2x_4 & x_1x_3^3 & x_1x_3^2x_4
 & x_1x_3x_4^2 & x_1x_4^3 & x_2^4 
 \\ x_2^3x_3 & x_2^2x_3^2 & x_2x_3x_4 
 & x_2^2x_4^2 & x_2x_3^3 & x_2x_3^2x_4 
 & x_2x_4^3 & x_3^4 & x_3^3x_4 
 & x_3^2x_4^2 \\ & & & & & & & & x_3x_4^3 & x_4^4
 		]^T
\end{array}
\end{equation}
with the size of  $ L_V=42 $, actor NN activation function vector as
\begin{equation} \label{eq_6.5}
 \begin{array}{ccccc}
  \psi(x) = [
x_1 & x_2 & x_3 & x_4 & \varphi ^ T(x)
 		]^T
 \end{array}
\end{equation}
with the size of  $ L_u=46 $, and initial actor NN weight vector as 
\begin{equation} \label{eq_6.6}
 \begin{array}{ccccccc}
\theta_{u}^{(0)} = [
1.0 & 1.0 & -0.7 & -2.0 & 0 & ... & 0
 		]^T.
 \end{array}
\end{equation} 
With the initial control policy $ \widehat {u}^{(0)}  $ obtained based on the actor NN weight vector $ \theta_{u}^{(0)} $, closed-loop simulation is conducted. Figures \ref{fig4} and \ref{fig6} demonstrate the trajectories of control action and states. To show the real cost generated by a control policy $ u $, define
\begin{equation} \label{eq_6.7}
J(t) \triangleq \int_0^t Q(x(\tau)) + W(u(\tau))d \tau.
\end{equation} 
Figure \ref{fig5} gives the trajectory of $ J(t) $ by using initial control policy $ \widehat {u}^{(0)}  $, from which it is observed that $ J(t) $ approaches to 0.9162 as time increases. 

In order to collect sample $ \mathcal {S}_M $ and compute $ \rho _ {\Delta \varphi} (x_k), \rho _ Q (x_k), \rho _ \psi^{l} (x_k) $ and $ \rho _ {u \psi}^{l} (x_k,u_k)$, let sample size  $ M = 1001 $ and time interval  $ \Delta t = 0.1s $. Then, we conducted closed-loop simulation on system \eqref{eq_6.3} with input signal $ u = \widehat {u}^{(0)} +e_u $, with $ e_u $ generated by \eqref{eq_6.2} that is shown in Figure \ref{fig7}. After the online procedure (i.e., Step 1) is completed, offline iteration (i.e., Steps 2-4) is employed to learn the optimal control policy. Setting the value of convergence criterion  $ \xi = 10^{-5} $,  it is indicated that the critic NN weight vector converges at the $8^{th}$ iteration to
\begin{figure}
	\begin{minipage}[t]{0.5 \linewidth}
	\setcaptionwidth{2.5in}
		\centering	\includegraphics[width=2.5in]{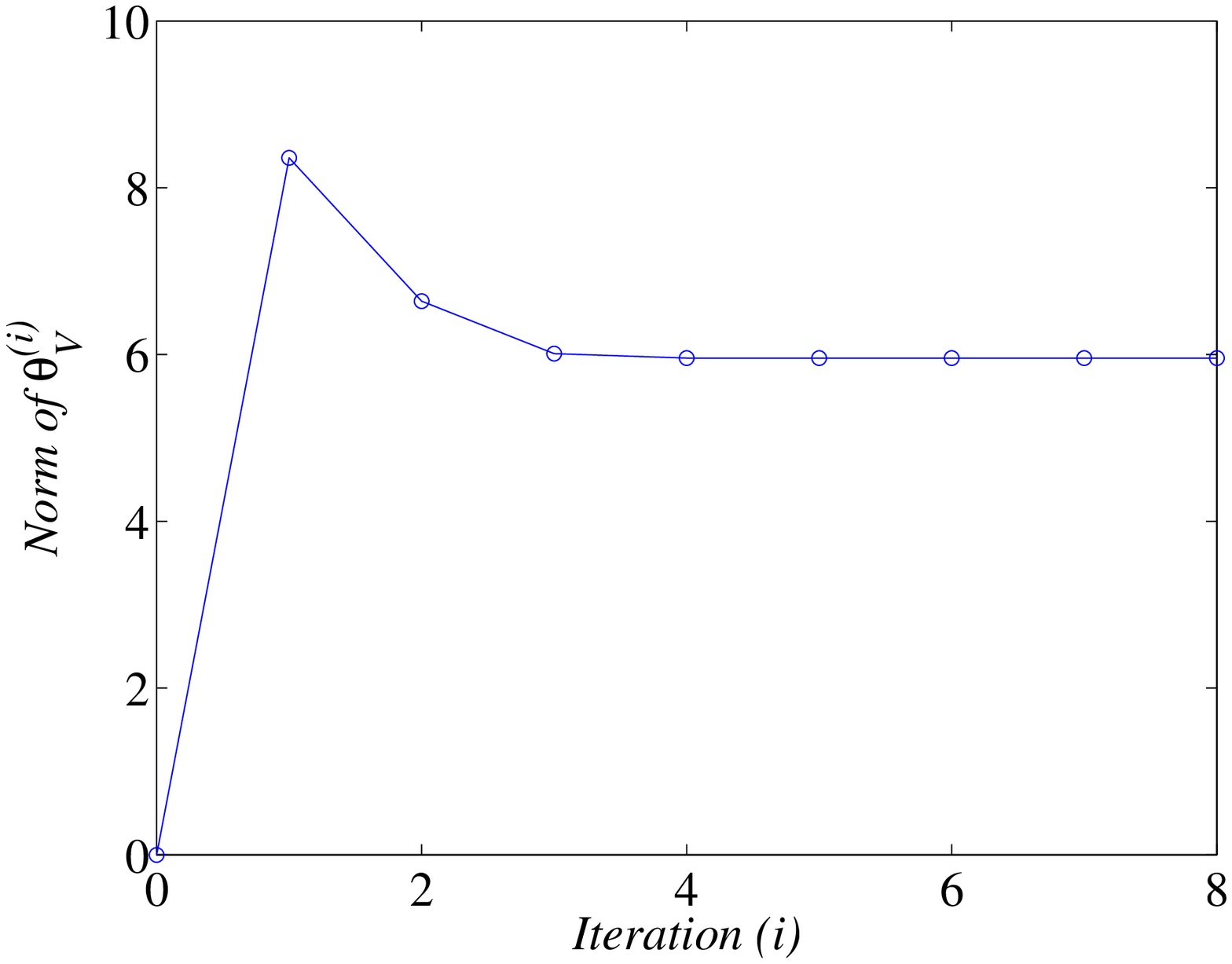}
	\caption{For case 2, the norm of critic NN weight vector $ \Vert \theta_{V}^{(i)} \Vert$  at each iteration.}
		\label{fig10}
	\end{minipage}%
	\begin{minipage}[t]{0.5\linewidth}
	\setcaptionwidth{2.5in}
		\centering	\includegraphics[width=2.5in]{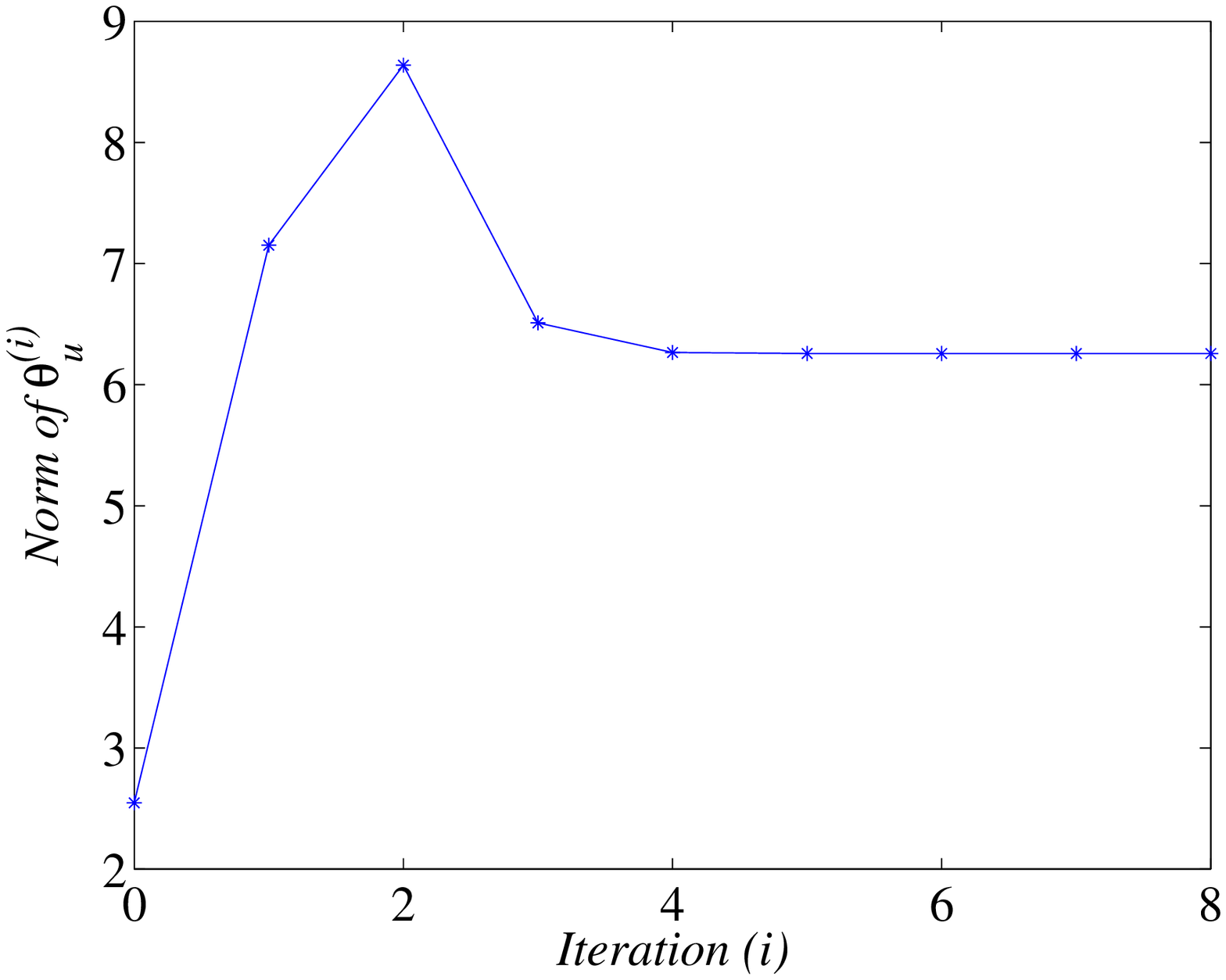}
		\caption{For case 2, the norm of actor NN weight vector $ \Vert \theta_{u}^{(i)} \Vert$ at each iteration.}
		\label{fig11}
	\end{minipage}
\end{figure}

\begin{figure}
	\begin{minipage}[t]{0.5 \linewidth}
	\setcaptionwidth{2.5in}
		\centering	\includegraphics[width=2.5in]{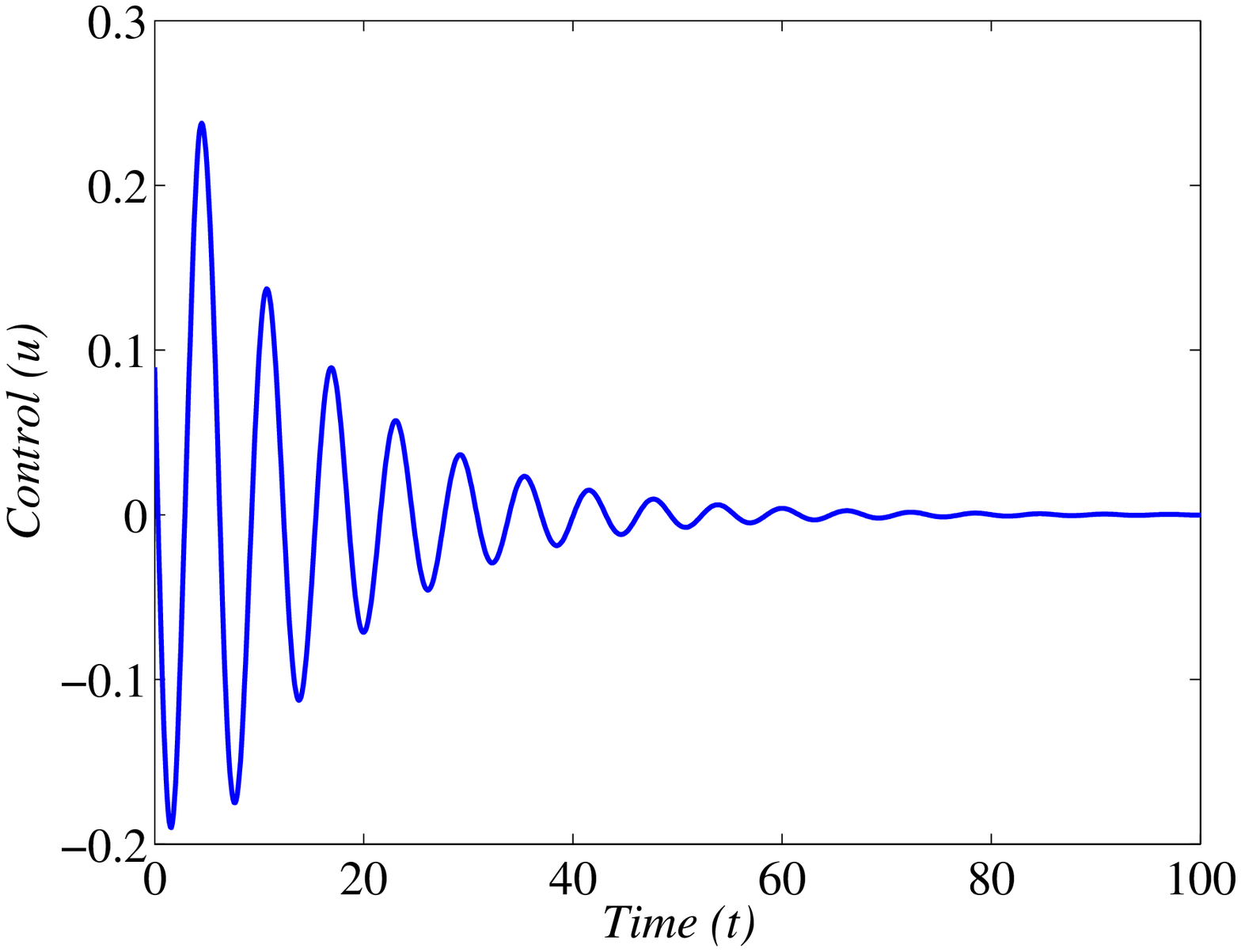}
	\caption{For case 2, trajectory of the final control policy $ \widehat {u}^{(8)} $.}
		\label{fig12}
	\end{minipage}%
	\begin{minipage}[t]{0.5\linewidth}
	\setcaptionwidth{2.5in}
		\centering	\includegraphics[width=2.5in]{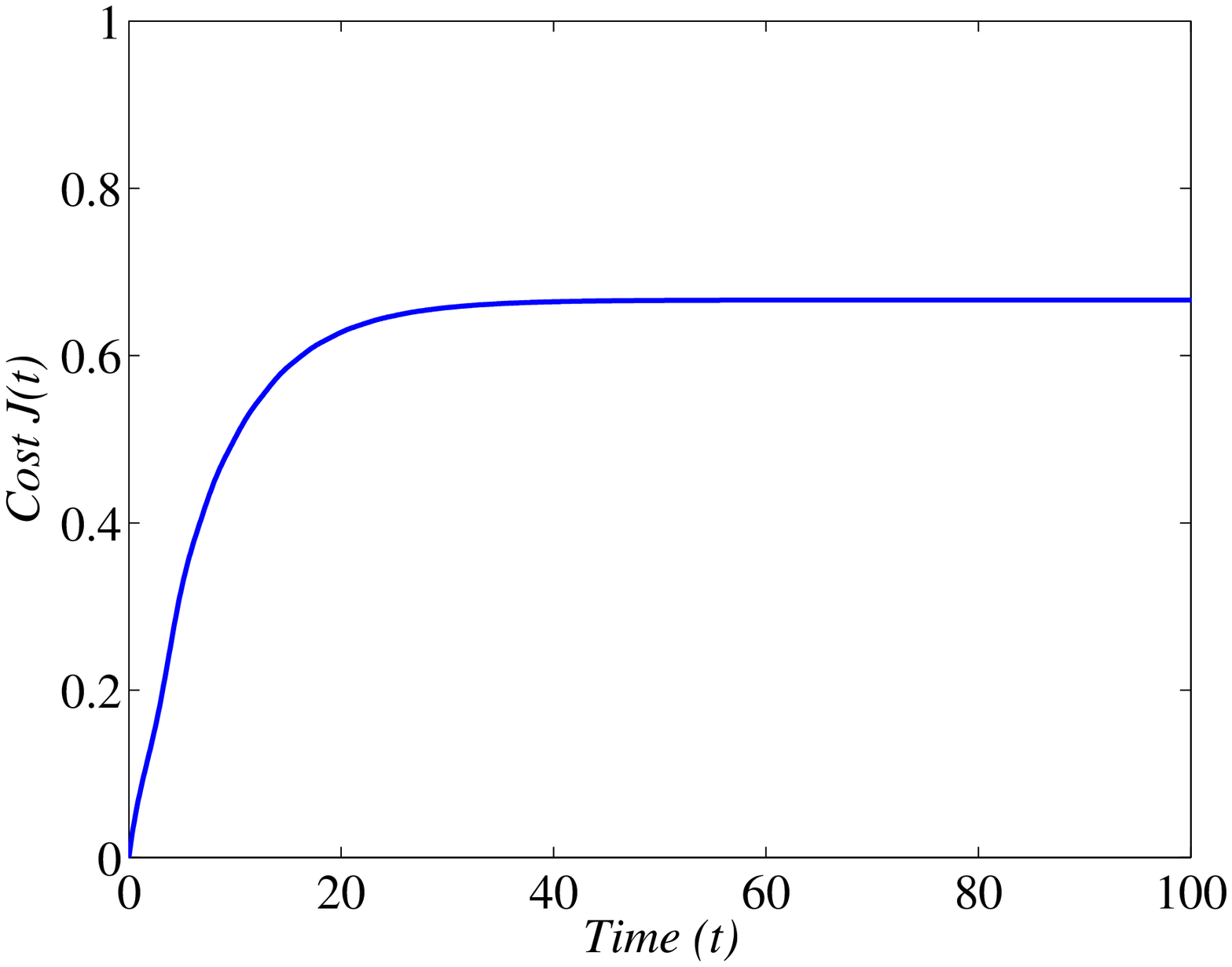}
		\caption{For case 2, trajectory of cost $J(t)$ with the final control policy $ \widehat {u}^{(8)} $.}
		\label{fig13}
	\end{minipage}
\end{figure}
\begin{equation}
 \begin{array}{ccccccccc}
\theta_{V}^{(8)} = [
       4.1255 & 0.0986 & -0.1745 & -0.8646 &  3.8691 &  0.2985
   &  0.8022 &  0.1645 &  0.4876 \\  0.7840 &  0.1230 &  0.4368 
   &  0.1788 &  0.1742 & -0.1535 &  0.0862 & -0.4266 & -0.2029 
   \\ -0.1242 &  0.0819 &  0.0922 & -0.0243 &  0.0433 & -0.0045
   & -0.0243 &  0.1474 &  0.1173 \\  0.0679 &  0.0384 &  0.1120
   &  0.0576 &  0.1803 &  0.0028 &  0.1211 &  0.1005 & -0.6450
   \\ & & & -0.0114 & -0.0574 & -0.0070 & -0.0923 & -0.0364 & -0.0088
 		]^T
\end{array} \nonumber
\end{equation} 
and the actor NN weight vector converges to 
\begin{equation}
 \begin{array}{ccccccccc}
\theta_{u}^{(8)} = [
    0.4602 & 0.3880 & -0.2227 & -0.7329 & 0.0032 & -0.0000
   & -0.0071 & -0.0050 & 0.0037 \\ 0.0010 & -0.0025 & 0.0013
   & 0.0028 & 0.0013 & -0.6310 & -0.1456 & 0.7415 & 0.2604
   \\ 1.9839 & 0.4553 & 0.5933 & -2.3179 & -1.3054 & 0.4513
   & 0.5094 & 0.0655 & -1.6990 \\ -0.7827 & 0.0655 & -0.9421
   &  2.1979 & 2.6326 & 0.4984 & -0.3965 & -1.6696 & -1.9667
   \\ 0.1059 & -0.3985 & -0.2620 & 0.5763 & 0.2819 & 0.2968
   & -0.3604 & -0.9616 & -0.3259 \\
   & & & & & & & &  -0.0005 \nonumber
 		]^T.
\end{array}
\end{equation}

\begin{figure}[htbp]
\centering\includegraphics[width=5in]{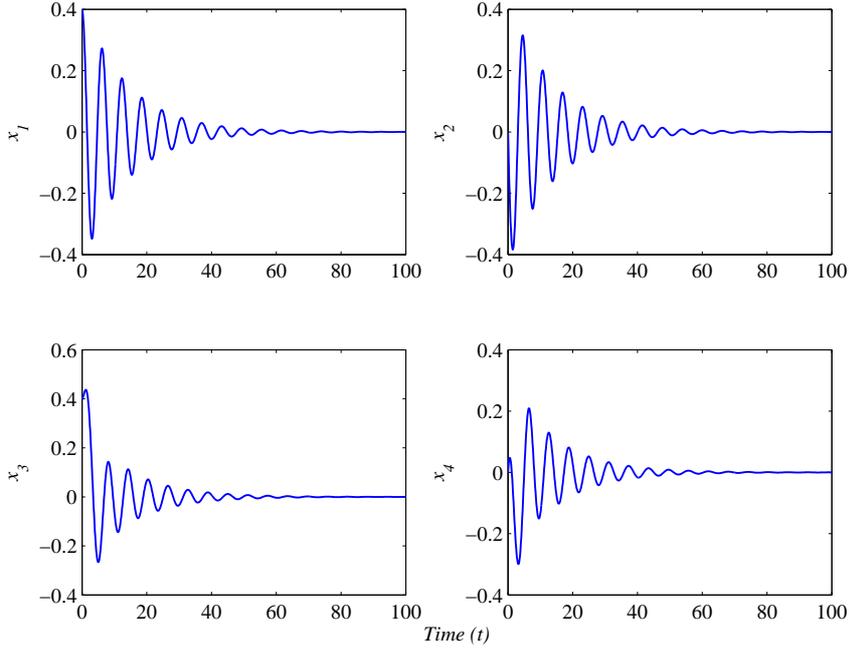}
\caption{For case 2, system state trajectories with the final control policy $ \widehat {u}^{(8)} $.}\label{fig14}
\end{figure}

Figures \ref{fig8} and \ref{fig9} show the first six representative critic NN weights $ \theta_{V,1}^{(i)}-\theta_{V,6}^{(i)} $, and the first six representative actor NN weights  $ \theta_{u,1}^{(i)}-\theta_{u,6}^{(i)} $ at each iteration. For brevity, we omit the figure of other elements of $ \theta_{V}^{(i)} $ and $ \theta_{u}^{(i)} $, and alternatively give the norm of critic and actor NN weight vectors (i.e., $ \Vert \theta_{V}^{(i)} \Vert$ and $ \Vert \theta_{u}^{(i)} \Vert$) in Figures \ref{fig10} and \ref{fig11}. It is shown from the figures that $ \Vert \theta_{V}^{(i)} \Vert$ and $ \Vert \theta_{u}^{(i)} \Vert$ converge to 5.9573 and 6.2560 respectively. By using the convergent actor NN weights  $ \theta_{u}^{(8)} $, closed-loop simulation is conducted with final control policy $ \widehat {u}^{(8)} $.  Figures \ref{fig12} and \ref{fig14} demonstrate the control action and state trajectories, respectively. The real cost $ J(t) $ is computed and shown in Figure \ref{fig13}, where $ J(t) $ converges 0.6665 to as time increases. This means that compared with initial control policy $ \widehat {u}^{(0)} $, the final control policy $ \widehat {u}^{(8)} $ obtained by the data-based API algorithm can reduce 27.42\%  of the cost (i.e., $ 1-0.6665/0.9162 = 0.2742 $). 

\begin{figure}[htbp]
\centering\includegraphics[width=5.0in]{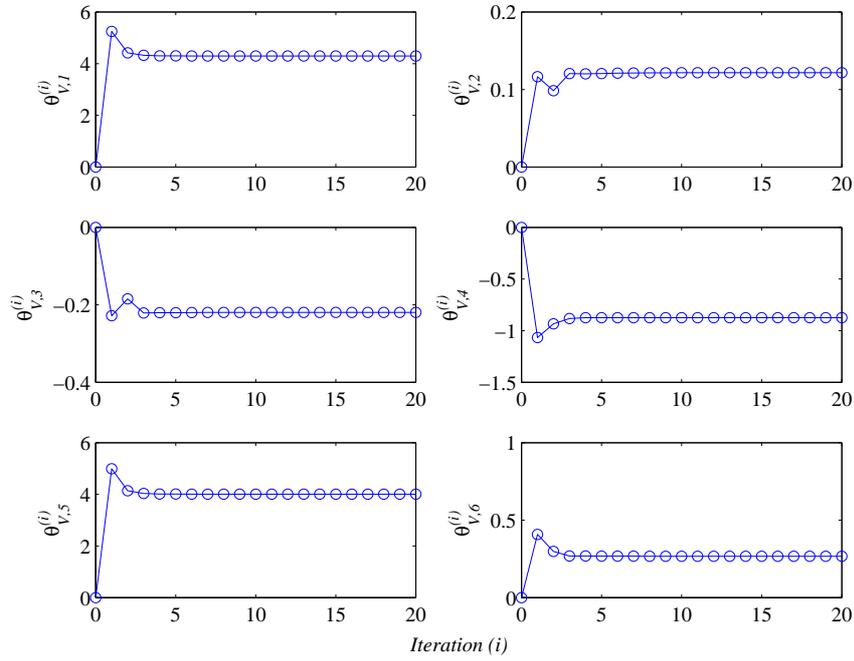}
\caption{For case 3, the first six representative critic NN weights $ \theta_{V,1}^{(i)}-\theta_{V,6}^{(i)} $ at each iteration.} \label{fig15}
\end{figure}

\begin{figure}[htbp]
\centering\includegraphics[width=5.0in]{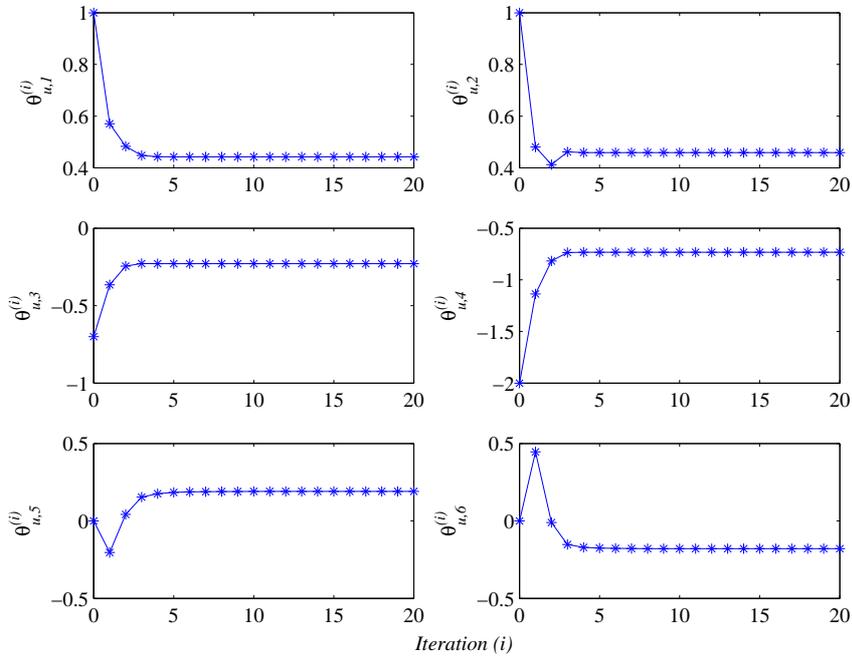}
\caption{For case 3, the first six representative actor NN weights $ \theta_{u,1}^{(i)}-\theta_{u,6}^{(i)} $ at each iteration. }\label{fig16}
\end{figure}

\subsection{Case 3: Application to the constrained RTAC nonlinear benchmark problem} \label{Sec_6.3}
Consider the constrained optimal control problem of the RTAC nonlinear benchmark problem given in Subsection \ref{Sec_6.2}, with the input constraint $ | u | \leqslant \beta,  \beta = 0.2$. Select $ \phi(\mu)  = \beta \tanh(\mu / \beta) $ and $ R  = 1$, then $ W(u) $ in cost functional \eqref{eq_2.2} is 
\begin{flalign}
W(u) 
&= 2\int_{0}^{u} \beta \tanh^{-1}(\mu / \beta) R d\mu \nonumber \\
&=2 \beta R u \tanh^{-1}(u / \beta) + \beta^2 R \ln (1-u^2/\beta^2). \nonumber
\end{flalign}

From Figures \ref{fig4} and \ref{fig12} associated with the initial and final unconstrained control policies in above Subsection \ref{Sec_6.2}, it is found that both control actions violate the constraint $ \beta $. To solve the constrained optimal control problem of system \eqref{eq_6.3} with the developed data-based API algorithm (i.e., Algorithm \ref{algorithm_4.1}), we choose the same critic NN activation function vector \eqref{eq_6.4}, actor NN activation function vector \eqref{eq_6.5} and initial actor NN weight vector \eqref{eq_6.6}. Using the exploratory noise $ e_u $  generated by \eqref{eq_6.2},  closed-loop simulation is conducted with $ u = \phi(\nu), \nu = \widehat {\nu}^{(0)} +e_u $. Then, collect sample set $ \mathcal{S}_M $ with size  $ M = 1001 $ and time interval  $ \Delta t = 0.1s $,  and compute $ \rho _ {\Delta \varphi} (x_k), \rho _ Q (x_k), \rho _ {u \psi}^{l} (x_k,u_k) $.  Setting $ \xi = 10^{-5} $, the simulation results show that, at the $20^{th}$ iteration, the critic and actor NN weight vector converge respectively to

\begin{figure}
	\begin{minipage}[t]{0.5 \linewidth}
	\setcaptionwidth{2.5in}
		\centering	\includegraphics[width=2.5in]{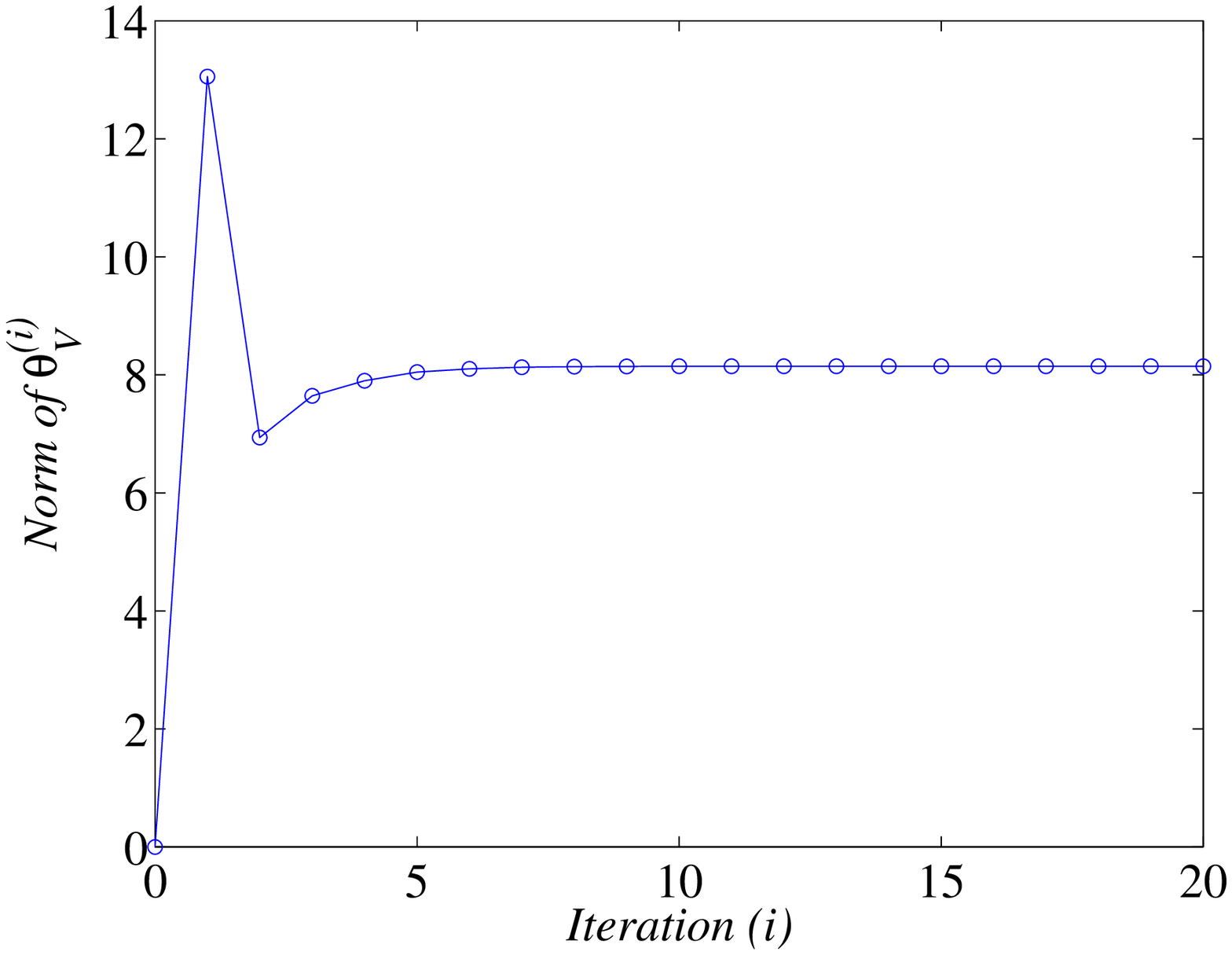}
	\caption{For case 3, the norm of critic NN weight vector $ \Vert \theta_{V}^{(i)} \Vert$  at each iteration.}
		\label{fig17}
	\end{minipage}%
	\begin{minipage}[t]{0.5\linewidth}
	\setcaptionwidth{2.5in}
		\centering	\includegraphics[width=2.5in]{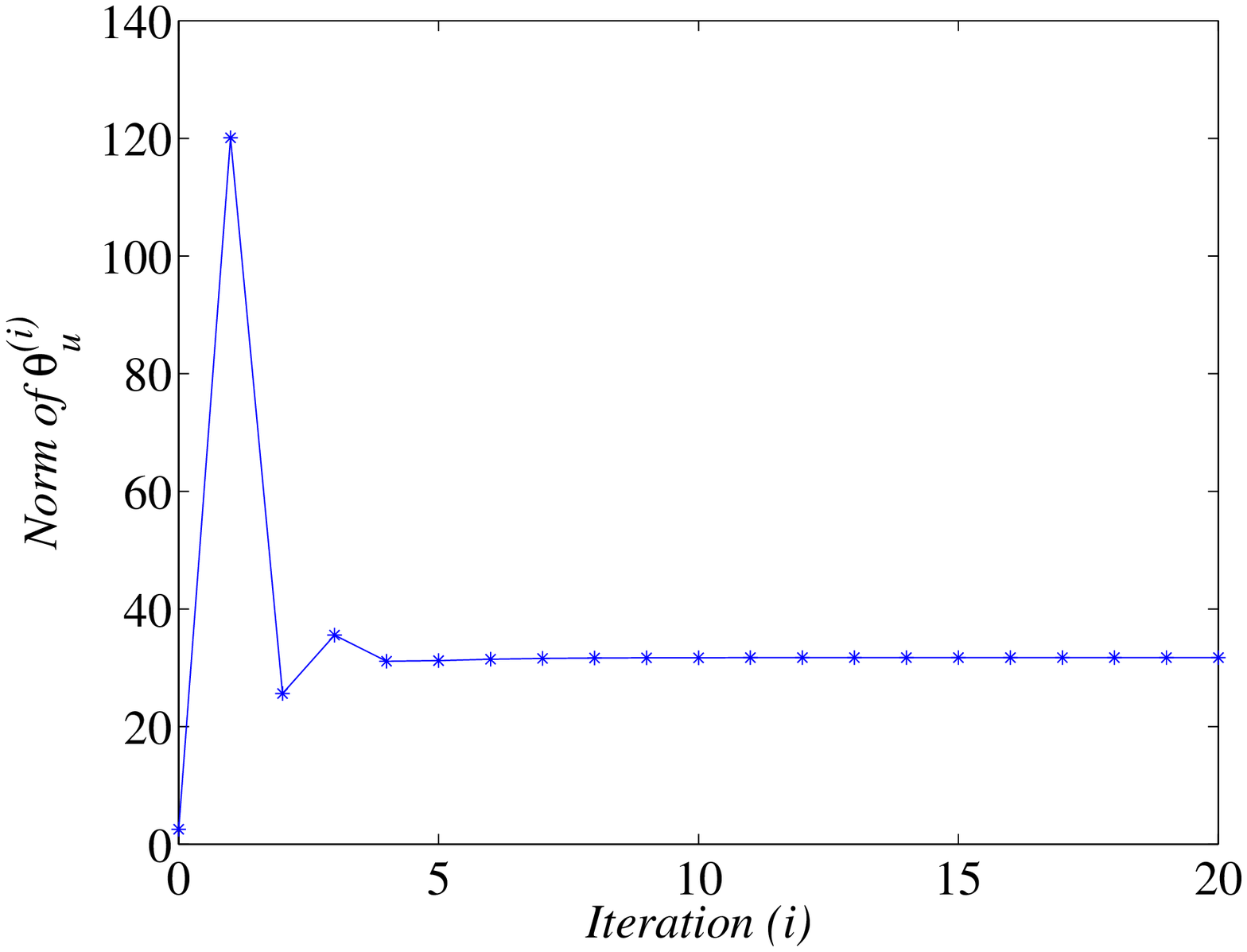}
		\caption{For case 3, the norm of actor NN weight vector $ \Vert \theta_{u}^{(i)} \Vert$  at each iteration.}
		\label{fig18}
	\end{minipage}
\end{figure}

\begin{figure}
	\begin{minipage}[t]{0.5 \linewidth}
	\setcaptionwidth{2.5in}
		\centering	\includegraphics[width=2.5in]{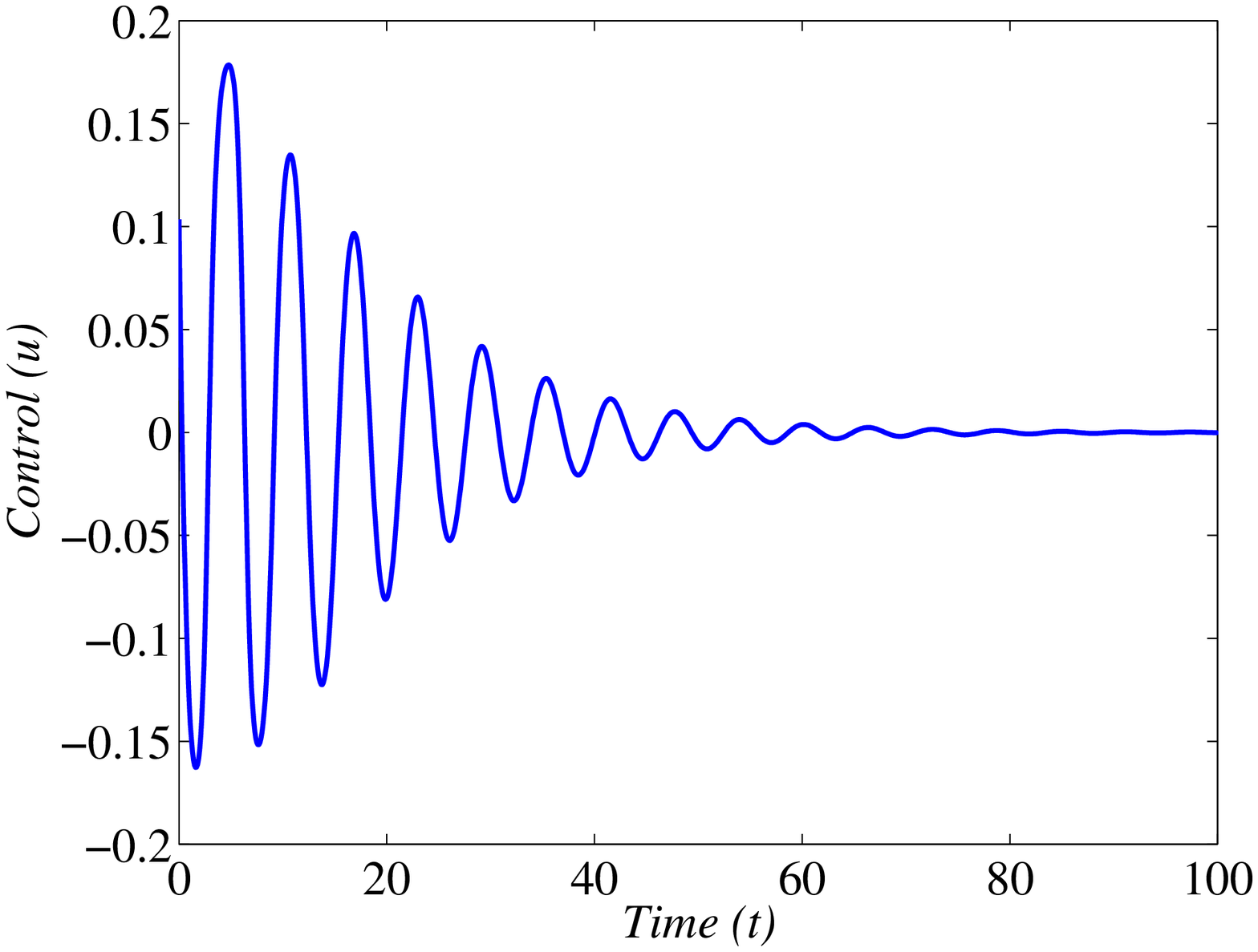}
	\caption{For case 3, trajectory of  the final control policy $ \widehat {u}^{(20)} $.}
		\label{fig19}
	\end{minipage}%
	\begin{minipage}[t]{0.5\linewidth}
	\setcaptionwidth{2.5in}
		\centering	\includegraphics[width=2.5in]{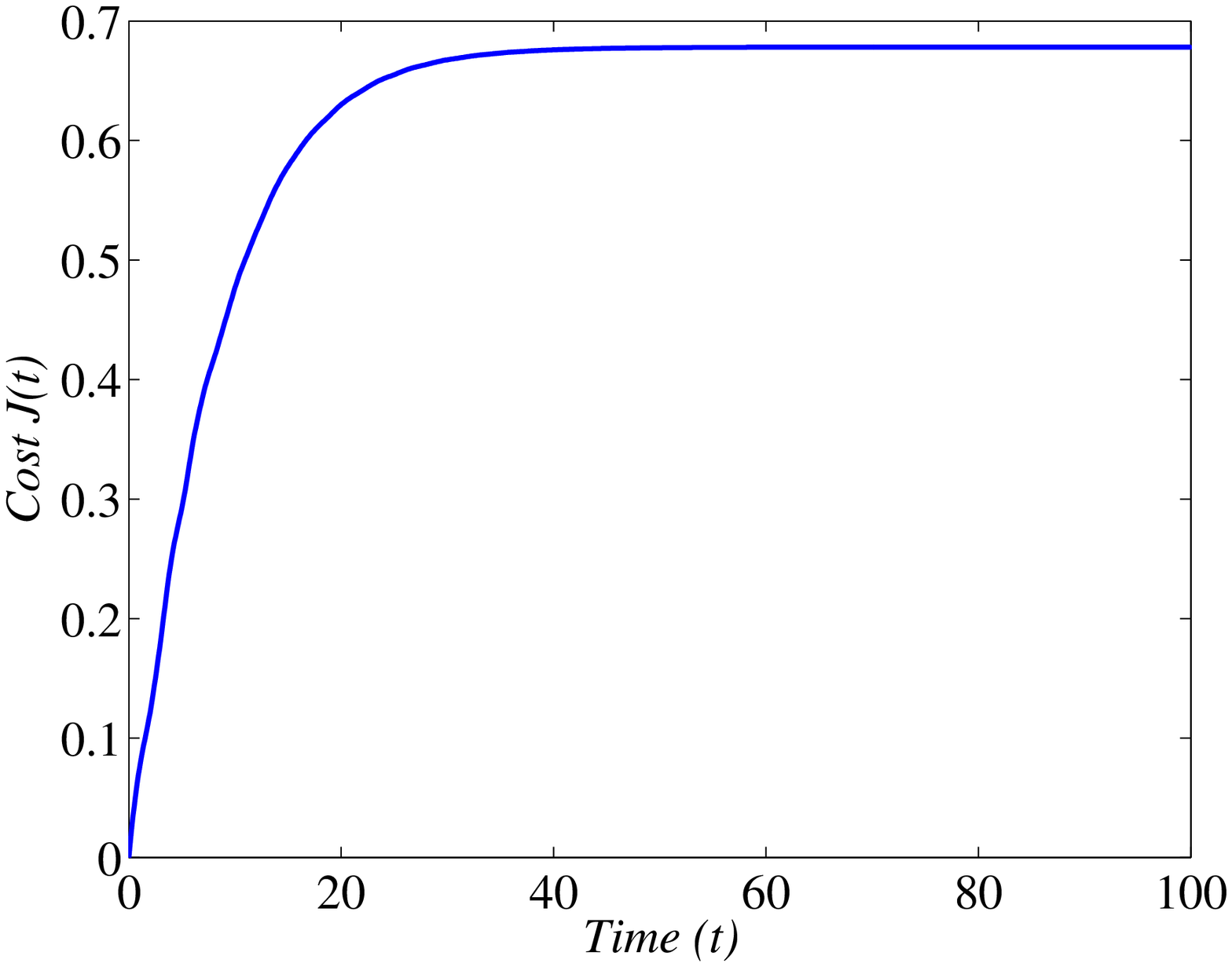}
		\caption{For case 3, trajectory of cost $J(t)$  with the final control policy $ \widehat {u}^{(20)} $.}
		\label{fig20}
	\end{minipage}
\end{figure}
\begin{figure}[htbp]
\centering\includegraphics[width=5in]{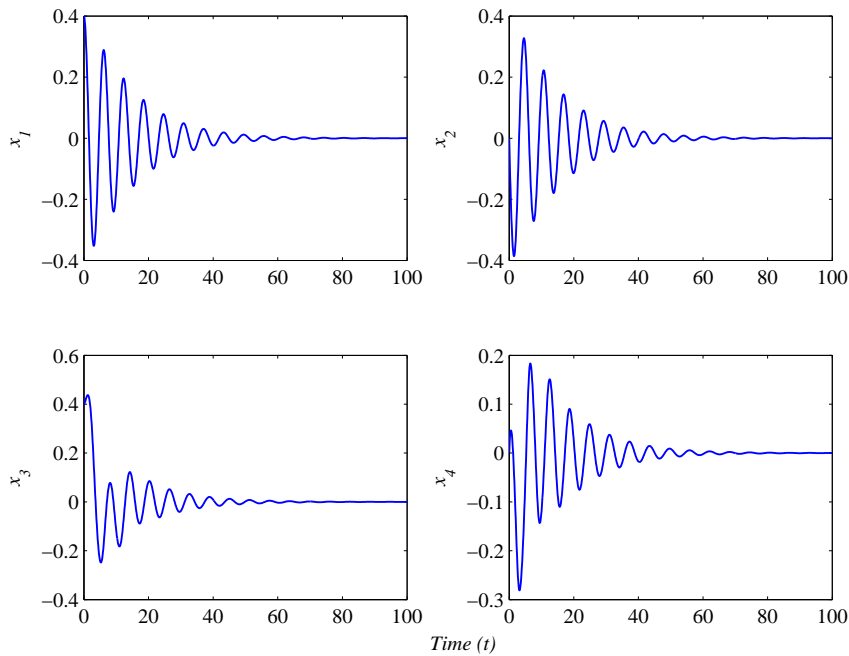}
\caption{For case 3, system state trajectories with the final control policy $ \widehat {u}^{(20)} $.}\label{fig21}
\end{figure}


\begin{equation}
 \begin{array}{ccccccccc}
\theta_{V}^{(20)} = [
    4.2970 & 0.1216 & -0.2196 & -0.8742 & 4.0075 & 0.2672 & 
    0.7472 & 0.1643 & 0.4953 \\ 0.7819 & 0.8625 & 1.9686 & 
    2.1268 & 0.5542 & -1.6487 & -0.3671 & -0.6932 & -2.4891 \\ 
   -0.9257 & 0.5616 & 1.4230 & 0.4928 & 0.6333 & 0.4141 & 
    0.4928 & 0.0537 & 0.1653 \\ 0.9117 & 1.4092 & 0.4641 & 
   -1.0433 & -0.1432 & 0.0280 & 1.0835 & 0.1701 & -0.5234 \\ 
   & & & -0.5237 & -0.0486 & -0.0376 & 0.2384 & -0.0521 & -0.5238
 		]^T
\end{array} \nonumber 
\end{equation}
\begin{equation}
 \begin{array}{ccccccccc}
\theta_{u}^{(20)} = [
    0.4421 & 0.4591 & -0.2291 & -0.7333 & 0.1905 & -0.1791 & 
   -0.0575 & -0.2978 & -0.0232 \\ -0.0739 & 0.1672 & -0.0035 & 
   -0.0009 & 0.0519 & 0.4666 & -4.6554 & -3.1500 & -0.9666 \\ 
    1.2378 & 0.2776 & 0.1788 & 8.9946 & 3.1199 & 3.2885 & 
    2.5886 & -3.9747 & 3.9289 \\ -18.3913 & -3.9747 & 4.6712 & 
   -5.3350 & -6.1230 & 14.1708 & -1.4422 & 0.7945 & 3.5809 \\ 
    0.7768 & 2.1957 & 1.9014 & -1.5518 & -7.0940 & 0.1421 & 
   -0.4144 & -0.1659 & 0.3712 \\
  & & & & & & & &   -10.5140
 		]^T.
\end{array} \nonumber
\end{equation}
Figures \ref{fig15} and \ref{fig16} demonstrate the first six representative critic NN weights $ \theta_{V,1}^{(i)}-\theta_{V,6}^{(i)} $, and the first six representative actor NN weights  $ \theta_{u,1}^{(i)}-\theta_{u,6}^{(i)} $ at each iteration. The norm of critic and actor NN weight vectors are shown in Figures \ref{fig17} and \ref{fig18}, where $ \Vert \theta_{V}^{(i)} \Vert$ and $ \Vert \theta_{u}^{(i)} \Vert$ converge to 8.1462 and 31.7143 respectively. By using the convergent actor NN weight vector  $ \theta_{u}^{(20)} $, closed-loop simulation is conducted with the final control policy $ \widehat {u}^{(20)} $, and Figures \ref{fig19} and \ref{fig21} give the trajectories of control action and states, respectively. It is indicated from Figure \ref{fig19} that the control constraint $ | u | \leqslant 0.2$ is satisfied. The real cost $ J(t) $ is computed and shown in Figure \ref{fig20}, where $ J(t) $ converges 0.6781 to as time increases.  

\section{Conclusions} \label{Sec_7}
The model-free constrained and unconstrained optimal problems of nonlinear continuous-time systems is addressed by proposing a data-based API algorithm, and its convergence is proved. The data-based API method learns the solution of HJB equation and the optimal control policy from real system data instead of mathematical model. The implementation procedure of the algorithm is based on the actor-critic-NN structure, which contains an online part for system information collection, and an offline part for iterative learning the optimal critic and actor weight vectors. The application on a simple nonlinear numerical system and a RTAC benchmark system demonstrate the effectiveness of the developed data-based API optimal control design method.



\bibliographystyle{elsarticle-num} 
\bibliography{References}






\end{document}